\documentclass[a4paper,11pt]{article}      
\usepackage{jcappub} 
\usepackage{multirow}
\usepackage{color}    
\RequirePackage{graphicx}
\usepackage{amsmath}    
\usepackage{graphicx}
\usepackage{makecell}
\usepackage{subcaption}   
\usepackage{tabularx}

\title{\boldmath   A cosmographic analysis using DESI- DR2 and strong lensing: I. Time- Delay measurements 
                  }

\author[a]{Darshan Kumar,}
\author[b]{Deepak Jain,}
\author[c]{and Shobhit Mahajan,}    

\affiliation[a]{Institute for Gravitational Wave Astronomy, Henan Academy of Sciences, \\ Zhengzhou 450046, Henan, China}
\affiliation[b]{Deen Dayal Upadhyaya College, University of Delhi, \\ Dwarka, New Delhi 110078, India}
\affiliation[c]{Department of Physics and Astrophysics, University of Delhi, \\ Delhi 110007, India}

\emailAdd{kumardarshan@hnas.ac.cn}
\emailAdd{djain@ddu.du.ac.in}  
\emailAdd{sm@physics.du.ac.in}  

\abstract{    
Strong gravitational lensing time-delay measurements, together with the distance sum rule (DSR), offer a model-independent approach to probe the geometry and expansion of the universe without relying on a fiducial cosmological model. In this work, we perform a cosmographic analysis by combining the latest Type Ia supernova datasets (PantheonPlus, DESY5, and Union3), baryon acoustic oscillation data from DESI-DR2, and updated time-delay distances from strong lensing systems. 
The analyses using SGL with individual SNIa datasets (SGL+PantheonPlus, SGL+DESY5, and SGL+Union3) indicate a preference for an open universe, though they remain consistent with spatially flat universe at the $95\%$ confidence level. When DESI-DR2 data is included in each combination, the constraints tighten and shift slightly toward a closed universe, while flatness remains supported at the $68\%$ confidence level. The best-fit values of $q_0$ and $j_0$ agree with $\Lambda$CDM expectations within $95\%$ or $99\%$ confidence depending on the dataset, whereas $s_0$ remains weakly constrained in all cases. 
{This work is the first in a series of two companion papers on cosmography with DESI-DR2 and strong lensing.} 

\vspace{2mm}
}     

\begin{document}
\maketitle  
\flushbottom

\section{Introduction}\label{sec_intro}     
The standard spatially flat $\Lambda$CDM model, which includes a cosmological constant and cold dark matter, explains most current observations but still faces unresolved issues such as the Hubble tension, spatial curvature inconsistencies, and the long-standing fine-tuning and coincidence problems \cite{1989RvMP...61....1W,1999PhRvL..82..896Z,2000astro.ph..5265W,2001PhRvL..87n1302D,2010ApJ...711..439C,2011MNRAS.416.1099C}. In particular, a significant $5\sigma$ tension exists between the value of the Hubble constant inferred from CMB anisotropies within $\Lambda$CDM \cite{2020A&A...641A...6P} and the direct measurements based on the Cepheid-calibrated distance ladder by the SH0ES collaboration \cite{2022ApJ...934L...7R}. This discrepancy may reflect unknown systematics or the presence of new physics beyond the standard model \cite{2017NatAs...1E.169F,2021APh...13102605D}. Moreover, different analyses of Planck data have pointed toward a closed universe with non-zero curvature \cite{2020NatAs...4..196D,2021PhRvD.103d1301H}, while the combination with low-redshift BAO measurements favors a flat universe with curvature parameter $\Omega_{k0} = 0.0007 \pm 0.0019$ \cite{2020A&A...641A...6P}, leaving the problem unsettled.\\   

A central challenge in this context is to determine whether the universe is open, flat, or closed, since the curvature affects both cosmic dynamics and the interpretation of observations. Model-dependent estimates often assume non-flat $\Lambda$CDM, which may bias the results \cite{2020NatAs...4..196D,2020MNRAS.496L..91E}. To overcome this, several model-independent approaches have been suggested. Clarkson et al. \cite{2008PhRvL.101a1301C} proposed a curvature test based on derivatives of distance with respect to redshift, though the method suffers from large uncertainties. More recently, the distance sum rule has been introduced as a simple and powerful tool under the assumption of the FLRW metric \cite{2015PhRvL.115j1301R}. Any violation of this rule would indicate a breakdown of the metric itself, while consistency across different datasets provides robust constraints on $\Omega_{k0}$.  \\

In the past three decades, the development of advanced space- and ground-based telescopes for imaging and spectroscopic observations has facilitated the discovery of a significant number of strong gravitational lensing (SGL) systems. The available sample of SGL systems has now become sufficiently large to allow statistical analyses that can be used to study lens properties \cite{2008MNRAS.384..843M,2021MNRAS.503.1319G,2005ApJ...630..764C}, distribution of dark matter \cite{1993ApJ...407...33M,2009ApJ...706.1078N,2012A&A...538A..43C,2021PhRvD.103f3511K,2022A&A...659L...5C} and put constraints on cosmological parameters \cite{2014ApJ...788L..35S,2017JCAP...03..028R,2017MNRAS.465.4914B,2019MNRAS.488.3745C}. Since the number of observed galaxy-scale SGL systems far exceeds that of cluster-scale systems, most statistical investigations have relied on the galaxy-scale sample. \\

One of the powerful techniques in strong lensing, known as time-delay cosmography (TDC) provides a way  to determine the Hubble constant ($H_0$)  independent of both the local distance ladder and probes linked to sound-horizon physics \cite{1964MNRAS.128..307R,2024SSRv..220...48B,2024SSRv..220...12S,2024SSRv..220...87S,2022LRR....25....6M}. Improvements in photometric accuracy have allowed the identification of several multiply-imaged quasars and supernovae \cite{2018MNRAS.481.1041T,2023MNRAS.520.3305L,2015Sci...347.1123K,2021NatAs...5.1118R}, together with highly precise measurements of time-delays \cite{2020A&A...642A.193M,DES:2017uif}. Notably, the H0LiCOW and SHARP collaborations examined six lenses and reported $H_0$ estimates with precision between $4.3\%$ and $9.1\%$ \cite{2010ApJ...711..201S,2014ApJ...788L..35S,2017MNRAS.465.4895W,2019MNRAS.484.4726B,2020MNRAS.498.1440R,2019MNRAS.490.1743C}. The STRIDES collaboration obtained an $H_0$ determination with $3.9\%$ precision by employing TDC as well \cite{2020MNRAS.494.6072S}. These studies generally rely on a standard framework that incorporates single-aperture stellar kinematics for each lens \cite{2017MNRAS.468.2590S}. The H0LiCOW team further constrained the value of $H_0 = 73.20 \pm 1.75$ km s$^{-1}$ Mpc$^{-1}$ with $2.4\%$ precision 
\cite{2020MNRAS.498.1420W}, which improved to $2\%$ once blind analyses were included \cite{2020A&A...639A.101M}. Further, for estimating $H_0$ and other cosmological parameters, the time-delay samples are combined with complementary distance indicators through the distance sum rule \cite{2015PhRvL.115j1301R}, an approach applied with type Ia supernovae (SNIa) \cite{2019PhRvL.123w1101C}, gamma-ray bursts (GRBs) \cite{2023MNRAS.521.4963D}, multi-messenger gravitational-wave standard sirens \cite{2019PhRvD..99h3514L}, ultra-compact radio sources \cite{2021MNRAS.503.2179Q}, quasars (QSOs) \cite{2020ApJ...897..127W}, and related probes. More recently, an inverse distance ladder based on the cosmic distance-duality relation (CDDR) \cite{Gong:2024yne,2023PhRvD.107b3520R} provided $H_0$ constraints consistent with SH0ES within $\sim 1.3\sigma$ \cite{Gong:2024yne}. Additionally, Gaussian process reconstructions using quasars and time-delay distances produced $H_0 = 70.8 \pm 1.5~ \mathrm{km s}~^{-1}~ \mathrm{Mpc}~^{-1}$ with a $5\%$ distance uncertainty \cite{2024ApJ...960..103L}, in agreement with several other analyses adopting CDDR across diverse observational data \cite{2024JCAP...05..098C,2024PhLB..85738982G,2020ApJ...895L..29L,2025Univ...11...89C}.\\

The use of time-delay cosmography alone is not sufficient to achieve precise constraints on the Hubble constant, cosmic curvature or other cosmological parameters. This limitation arises because lensing distances must be anchored with other reliable distance indicators to avoid relative calibration issues. Supernovae, gamma-ray bursts, quasars, and other probes have been combined with lensing systems, but the need for such external calibrations naturally introduces additional sources of uncertainty. Moreover, depending on specific background models can introduce circularity, as the calibration may be affected by the very cosmology under investigation. These challenges motivate the importance of alternative frameworks where cosmic distances can be reconstructed without assuming a specific cosmological model or secondary observations.\\

A promising route in this regard is the cosmographic approach, which focuses on the kinematics of cosmic expansion rather than the underlying dynamics \cite{1972gcpa.book.....W,1976Natur.260..591H,1997Sci...276...88V}. The essential idea is to expand observable quantities, such as the scale factor or cosmological distance, as a Taylor series around the present epoch \cite{2005GReGr..37.1541V,2011MPLA...26.1459L,2013MPLA...2850080L}. The coefficients of these expansions carry physical meaning because they directly encode parameters such as the Hubble rate, the deceleration parameter, or higher-order derivatives that describe the evolution of the universe. Since this method only requires the assumption of large-scale homogeneity and isotropy, it avoids dependence on a specific cosmological model and thus circumvents the problem of circularity.\\

Accurate reconstruction of the universe's expansion history relies on cosmographic methods that are independent of specific cosmological models.  Among these, Taylor-based cosmography is widely used, but it faces a key limitation. One difficulty of Taylor-based cosmography is that the standard expansion in terms of redshift converges only at low redshift, typically for $z \leq 1$ \cite{1998tx19.conf.....P,2004JCAP...09..009C,2004ApJ...607..665R,2004CQGra..21.2603V}. The inability of this $z$-{dependent Taylor} 
expansion at high redshift has a strong impact on the resulting analysis \cite{2010JCAP...03..005V}. However, modern observational data extend well beyond this range, from supernovae at $z \approx 2.3$ up to the cosmic microwave background at $z \approx 1100$. To address the convergence issues at high redshifts, several improved redshift parametrizations have been proposed. These include reparameterizing the redshift variable through auxiliary definitions, such as the y-redshift \cite{2001IJMPD..10..213C,2003PhRvL..90i1301L} or $E(y)$ \cite{2020ApJ...900...70R}, and applying rational approximations to achieve a smooth evolution of observables, for example Padé \cite{2014PhRvD..89j3506G,2014JCAP...01..045W,2020MNRAS.494.2576C} and Chebyshev rational polynomials \cite{2012JCAP...05..024S,2018MNRAS.476.3924C}.\\

Motivated by the above points, in this analysis, we consider strong gravitational lensing time-delay distance measurements as a central probe to study cosmography up to the fourth order to explore the expansion history of the universe in a model-independent framework. In addition, we use the distance sum rule to study the curvature parameter, which provides an independent consistency check on the underlying cosmic geometry. The significance of gravitational lensing arises from its complementary sensitivity to the expansion rate and cosmic curvature and its role as an independent geometrical test of the universe. To further strengthen the analysis, we incorporate three recent and well-established samples of Type Ia supernovae—PantheonPlus, Union3, and DESY5—together with the latest BAO measurements from DESI-DR2 and explore the constraints on cosmic curvature parameters $(\Omega_{k0})$ and cosmgraphic parameters $(q_0,~j_0,~s_0)$. Such a comprehensive and complementary dataset provides a solid foundation for testing cosmological models using observations without assuming any specific background cosmology.\\


The paper is organized as follows. The observational datasets used in this work are described in Section \ref{sec_obse_data}. Section \ref{sec_meth} presents the methodology adopted in this analysis. The main results are reported in Section \ref{sec_resu}. Finally, the implications of these results and the conclusions of the study are discussed in Section \ref{sec_disc_conc}.

\section{Observational Datasets}\label{sec_obse_data}
In this section, we present the observational datasets used in our analysis. These include time-delay measurements from strong gravitational lensing, recent samples of Type Ia supernovae 
and baryon acoustic oscillations. Each dataset provides complementary information that constrains different aspects of the cosmological parameters.

\subsection{Strong Gravitational Lensing- Time-Delay Distance}
In strong gravitational lensing systems, time-delay serves as a crucial observable for constraining cosmological parameters in a model-independent manner. Light rays that originate simultaneously from the source reach the observer at different moments because the corresponding trajectories differ in path length and traverse distinct gravitational potentials. As a result, multiple images show measurable time-delays. When the source is intrinsically variable, these delays can be determined by monitoring the brightness variations of the lensed images, since they correspond to the same physical event. The concept of time-delay distance is introduced in this context, which directly connects to the determination of cosmological parameters.This distance depends on the combination of three angular diameter distances: the observer–lens distance $d_\mathrm{A}^{^\mathrm{ol}}$, the lens–source distance $d_\mathrm{A}^{^\mathrm{ls}}$, and the observer–source distance $d_\mathrm{A}^{^\mathrm{os}}$. \\

For a specific source position $\mathcal{B}$ and image position $\theta$, the time-delay difference $\Delta t$ between two images, say image $i$ and $j$, is  expressed as \cite{1990CQGra...7.1319P,1990CQGra...7.1849P}:

\begin{equation}\label{equ_time_diff}
\Delta t_{i j}\equiv \delta t_j-\delta t_i=\dfrac{\left(1+z_l\right)}{c}\frac{{d_\mathrm{A}^{^\mathrm{o s}} d_\mathrm{A}^{^\mathrm{o l}}}}{{d_\mathrm{A}^{^\mathrm{l s}}}}\left[\dfrac{\left({\theta}_{j}-{\mathcal{B}}\right)^{2}}{2}-\psi\left({\theta}_{j}\right)-\dfrac{\left({\theta}_{i}-{\mathcal{B}}\right)^{2}}{2}+\psi\left({\theta}_{i}\right)\right]
\end{equation}
where $z_l$ is the lens redshift and $\psi$ is the effective gravitational potential of the lens.\\

From Equation \ref{equ_time_diff}, we define the ``time-delay distance'', $d_{\Delta t}^{^{\mathrm {}}}$, as

\begin{equation}\label{equ_tdd}
    d_{\Delta t}^{^{\mathrm { }}} = \left(1+z_l\right)\frac{{d_\mathrm{A}^{^\mathrm{o s}} d_\mathrm{A}^{^\mathrm{o l}}}}{{d_\mathrm{A}^{^\mathrm{l s}}}}
\end{equation}

The determination of the time-delay distance in strong lensing systems requires three essential ingredients: an accurate measurement of the time-delay between lensed images, a detailed treatment of the lensing contributions from intervening structures along the line of sight, and reliable models for the lens galaxies. The lens galaxy potential, described through the mass model, governs the observed time-delays. However, matter distributed along the line of sight can also alter the trajectories of light rays, producing focusing or defocusing effects that lead to systematic shifts in the inferred time-delay distance \cite{1994ApJ...436..509S,2016A&ARv..24...11T}. This contribution is often parameterized by the external convergence, $\kappa_{\mathrm{ext}}$, which modifies both $d_{\Delta t}$ and the Hubble constant, $H_0$. Since the average of $\kappa_{\mathrm{ext}}$ vanishes across the sky, neglecting this effect in individual systems can still bias the estimates of $d_{\Delta t}$ and $H_0$. Moreover, lens modeling alone cannot fully determine $\kappa_{\mathrm{ext}}$ due to the mass-sheet degeneracy, making independent line-of-sight structure analyses necessary to obtain robust constraints \cite{2003ApJ...584..664K,2014MNRAS.443.3631M}. \\

In this analysis, we consider the sample taken from H0LiCOW ($H_0$ Lenses in COSMOGRAIL's Wellspring) collaboration, which analyzed six lenses i.e., B1608+656 \cite{2010ApJ...711..201S,2019Sci...365.1134J}, RXJ1131-1231 \cite{2013ApJ...766...70S,2014ApJ...788L..35S,2019MNRAS.490.1743C}, HE 0435-1223 \cite{2019MNRAS.490.1743C,2017MNRAS.465.4895W}, SDSS 1206+4332 \cite{2019MNRAS.484.4726B}, WFI2033-4723 \cite{2020MNRAS.498.1440R}, and PG 1115+080 \cite{2019MNRAS.490.1743C}. For all lenses except B1608+656, the data analysis was carried out without prior knowledge of the cosmological parameter values. The likelihood functions for the time-delay distances of the six lens systems can be accessed from the H0LiCOW website\footnote{\url{https://www.h0licow.org}}. Specifically, for the lens B1608+656, the likelihood is modeled using a skewed log-normal distribution, expressed as:
\begin{equation}\label{equ_likel_b1608}
\mathcal{L}_{D_{\Delta t}} = \frac{1}{\sqrt{2\pi} (x - \lambda_D) \sigma_D} \exp\left[-\frac{\left(\ln(x - \lambda_D) - \mu_D \right)^2}{2\sigma_D^2}\right],
\end{equation}
where the parameters are set to $\mu_D = 7.0531$, $\sigma_D = 0.22824$, and $\lambda_D = 4000.0$, with the dimensionless variable defined as $x = D_{\Delta t}/(1~\mathrm{Mpc})$. The time-delay distance for lensed quasar systems are reported in Table~2 of Ref.~\cite{2020MNRAS.498.1420W}.  Further, the STRIDES collaboration reported the most precise measurement of $H_0$ to date based on the single time-delay lens DES J0408$-$5354 \cite{2020MNRAS.494.6072S}. \\

Following the approach adopted by Collett et al.~\cite{2019PhRvL.123w1101C}, we also consider constraints derived from the double-source-plane strong lens system SDSSJ0946+1006 \cite{2008ApJ...677.1046G}. In this system, the lensing galaxy is located at a redshift of $z_l = 0.222$, while the first source, denoted as $s_1$, is at $z_{s1} = 0.609$ \cite{2008ApJ...677.1046G}. The redshift of the second source, $s_2$, is set to $z_{s2} = 2.3$, corresponding to the peak value of the photometric redshift probability distribution reported by Collett and Auger~\cite{2014MNRAS.443..969C}. The presence of two background sources lensed by the same foreground galaxy provides a robust geometric constraint through the cosmological scaling factor, defined as

\begin{equation}\label{equ_likel_beta}
\beta \equiv \frac{d_\mathrm{A}^{^\mathrm{ls1}} \, d_\mathrm{A}^{^\mathrm{os2}}}{d_\mathrm{A}^{^\mathrm{os1}} \, d_\mathrm{A}^{^\mathrm{ls2}}}=    \frac{d_\mathrm{A}^{^\mathrm{ls1}} }{d_\mathrm{A}^{^\mathrm{ol}} \, d_\mathrm{A}^{^\mathrm{os1}}}\times\frac{d_\mathrm{A}^{^\mathrm{ol}} \, d_\mathrm{A}^{^\mathrm{os2}}}{ d_\mathrm{A}^{^\mathrm{ls2}}}=\dfrac{d_{\Delta t}^{\mathrm{s_2}}}{d_{\Delta t}^{s_1}}
\end{equation}
where $d_\mathrm{A}^{^\mathrm{ls1}}$ and $d_\mathrm{A}^{^\mathrm{ls2}}$ represent the angular diameter distances between the lens and sources $s_1$ and $s_2$, respectively, and $d_\mathrm{A}^{^\mathrm{os1}}$ and $d_\mathrm{A}^{^\mathrm{os2}}$ are the angular diameter distances from observer to the sources. This ratio is particularly sensitive to the curvature parameter $\Omega_{k0}$ and remains independent of the Hubble constant $H_0$. For the SDSSJ0946+1006 system, the inverse scaling factor $\beta^{-1}$ is constrained to be $1.404 \pm 0.016$ \cite{2014MNRAS.443..969C}. The posterior distribution of $\beta^{-1}$ is well described by a Gaussian likelihood centered at this value, with standard deviation $\sigma_{\beta^{-1}} = 0.016$. The corresponding likelihood function takes the form
\begin{equation}\label{equ_likeli_beta}
\mathcal{L}_{\beta^{-1}} = \frac{1}{\sqrt{2 \pi \sigma_{\beta^{-1}}^2}} \exp \left[ - \frac{\left(\beta^{-1}_\mathrm{obs} - \beta^{-1}_\mathrm{th} \right)^2}{2 \sigma_{\beta^{-1}}^2} \right],
\end{equation}
where $\beta^{-1}_\mathrm{th}$ denotes the theoretical value of the dimensionless cosmological scaling factor. This likelihood provides a complementary constraint on $\Omega_{k0}$ and serves as an additional probe of the geometry of the universe.\\

In this analysis, we consider 8 datapoints for time-delay distance. 
{ The time-delay data samples consider combined} high-quality time-delay measurements from optical and radio monitoring, deep imaging with HST and adaptive optics, and spectroscopic observations of both lens galaxies and line-of-sight environments. To reduce uncertainties, only lenses with minimal contributions from time-delay measurements, line-of-sight effects, and lens modeling systematics were included. Thus, time-delay cosmography provides a powerful and independent method to constrain cosmological parameters.

\subsection{Baryon Acoustic Oscillations }           
Baryon Acoustic Oscillations (BAO) are regarded as the archetype of statistical standard rulers \cite{2010dken.book..246B}. They originate from oscillations in the photon–baryon plasma of the early universe, where radiation pressure and gravity acted in opposition until recombination allowed photons to decouple and form the CMB. The residual baryonic oscillations became imprinted as a preferred clustering scale, identified with the sound horizon \cite{1998ApJ...496..605E,1972CoASP...4..173S}. This scale appears in multiple observables, including the two-point correlation function, the galaxy power spectrum, and the CMB anisotropy spectrum \cite{1970ApJ...162..815P}, and it serves as a robust probe of the cosmic expansion history \cite{2007ApJ...664..675E,2012MNRAS.426.1280S}. \\

In this study, we use the second-year BAO dataset released by the Dark Energy Spectroscopic Instrument (DESI), which corresponds to three years of observations \citep{2025arXiv250314743D,2025arXiv250314738D}. The measurements incorporate the full covariance matrix in order to capture correlations among different distance indicators, thereby ensuring a consistent statistical framework. This makes the DESI BAO dataset particularly valuable for deriving constraints on cosmological parameters with improved precision.  \\

The observables extracted from BAO analysis are expressed in terms of three distance ratios, normalized by the sound horizon at the drag epoch $r_d$. These ratios are the transverse-comoving distance ratio $d_\mathrm{co}(z)/r_d$, the Hubble distance ratio $d_H(z)/r_d$, and the volume-averaged distance ratio $d_V(z)/r_d$, defined as follows:   

\begin{align} 
    \text{First quantity}: \quad & \dfrac{d_\mathrm{co}(z)}{r_d} = \dfrac{d_L(z)}{r_d\,(1+z)},  \label{eq_dM} \\
    \text{Second quantity}: \quad & \dfrac{d_H(z)}{r_d} = \dfrac{c}{r_d\, H(z)}, \label{eq_dH} \\
    \text{Third quantity}: \quad & \dfrac{d_V(z)}{r_d} = \dfrac{\left[z\, d_H(z)\, d_\mathrm{co}^2(z)\right]^{1/3}}{r_d}, \label{eq_dV}
\end{align}
The sound horizon $r_d$ depends on the physical matter and baryon energy densities as well as on the effective number of relativistic species, and therefore links BAO measurements to the physics of the early universe.  \\

It is important to highlight that BAO observables are sensitive to a strong degeneracy between the Hubble constant $H_0$ and the sound horizon $r_d$. As a result, BAO data alone cannot simultaneously determine both quantities. To address this issue, we adopt a value of the baryon density parameter, $\Omega_{b0}=0.02218 \pm 0.00055$, which is used to compute $r_d$ in our analysis~\cite{2025JCAP...02..021A}. This approach breaks the degeneracy and allows for meaningful cosmological constraints. As a result, the DESI-2025 BAO measurements provide an independent and complementary probe to other cosmological datasets and strengthen the robustness of parameter estimation within the $\Lambda$CDM framework and its possible extensions. In our analysis, we impose a Gaussian prior on $r_d$.

\subsection{Type Ia Supernova}
Type Ia supernovae (SNIa) are thought to be the result of the explosion of a carbon-oxygen white dwarf in a binary system as it goes over the Chandrashekhar limit, either due to accretion from a donor or mergers. They act as standard candles in determining cosmic distances through the use of the cosmic distance ladder. They are invaluable as they are bright enough to be observed at large cosmic distances, sufficiently common to be identified in significant numbers, and capable of being standardized. These objects were also responsible for providing the first evidence of an accelerating expansion phase of the universe \cite{1998AJ....116.1009R,1999ApJ...517..565P}.  \\

In this work, we consider three major compilations of SNIa: the PantheonPlus, Union3, and DESY5 samples. The PantheonPlus dataset is one of the most comprehensive collections of SNIa available. It consists of $1701$ light curves from $1550$ spectroscopically confirmed SNIa spanning the redshift range $0.001 \leq z \leq 2.261$ \cite{2022ApJ...938..113S}. 
The construction of this catalogue involves cross-calibration of photometric systems, forward modelling of supernova populations, and corrections for observational biases \cite{2021ApJ...909...26B,2021ApJ...913...49P,2023ApJ...945...84P}. The PantheonPlus public release provides the likelihood function with both statistical and systematic covariance matrices, and the dataset includes 203 spectroscopically confirmed DES year 3 supernovae \cite{2019ApJ...874..150B}. In cosmological analyses, a low-redshift cut of $z > 0.01$ is often imposed to reduce the impact of peculiar velocities on the Hubble diagram \cite{2022ApJ...938..112P}. Therefore, in our analysis, we exclude low redshift $(z \leq 0.01)$ samples and work with 1590 light curves. \\

The Union3 compilation provides an even larger sample of $2087$ SNIa \cite{2025ApJ...986..231R}, with $1363$ objects overlapping with PantheonPlus. Union3 adopts a different statistical methodology for the treatment of systematics, relying on Bayesian hierarchical modelling to account for measurement errors and observational biases. This approach makes Union3 particularly suitable for cross-validation against PantheonPlus, as well as for independent cosmological inferences. In this work, we use the latest Union~3.0 compilation of Type Ia supernovae observations spanning the redshift range $0.01 < z < 2.26$. It is important to note that, at present, only the binned distance modulus measurements are publicly available for this sample. Consequently, our analysis is restricted to these binned data points, and we also take into account their corresponding covariance matrix.\\   

The DESY5 dataset corresponds to the full five-year Dark Energy Survey Supernova Program and presents a homogeneous sample of $1635$ photometrically classified SNIa within $0.0596 < z < 1.12$, complemented by $194$ low-redshift SNIa at $z < 0.1$ from more recent surveys  \cite{2024ApJ...973L..14D}. Unlike PantheonPlus, DESY5 avoids the inclusion of older historical low-redshift samples to simplify the cross-calibration analysis. Most DES SNe lack spectroscopic confirmation, so a Bayesian classification algorithm (SuperNNova) is employed to assign probabilities for each SN to belong to the Ia population \cite{2017ApJ...836...56K,2020MNRAS.491.4277M}. These classification uncertainties are explicitly propagated into the covariance matrix. Importantly, the DESY5 dataset includes a larger fraction of high-redshift SNIa ($z > 0.5$) compared to PantheonPlus and thus is highly effective for probing the cosmic expansion history during epochs dominated by dark energy.\\

The PantheonPlus, Union3, and DESY5 compilations represent the most recent and comprehensive Type Ia supernova datasets, each characterized by distinct advantages in redshift coverage, statistical precision, and treatment of systematic uncertainties. In the present work, these datasets are analyzed separately and independently rather than in combination, in order to minimize cross-correlations and to examine the robustness of cosmological constraints obtained from different survey strategies. This approach preserves their individual features and obtains a direct test of consistency between the resulting parameter estimates.

\section{Methodology}\label{sec_meth}
In this work, we implement a modified framework of the Distance Sum Rule method to incorporate the information from time-delay distances measured in strong gravitational lensing systems. For this purpose, we use a cosmographic expansion up to the fourth order, which provides a model-independent approach to probe the history of the universe. The analysis combines multiple observational datasets, including the second data release of the Dark Energy Spectroscopic Instrument (DESI-DR2), three Type Ia supernova samples (PantheonPlus, Union3, and DESY5). 

\subsection{Cosmographic Expansions}   
{Cosmography has recently attracted growing interest, as it provides a way to extract information directly from observations under the minimal assumptions of isotropy and homogeneity. This approach does not invoke any specific dynamical model.} Within this approach, the only ingredient introduced a priori is the FLRW line element:

\begin{equation}\label{equ_flrw}
    ds^2=-c^2dt^2+a^2(t)\left[\dfrac{\\dr^2}{1-kr^2}+r^2d\Omega^2\right]
\end{equation}
where, $c$ is the speed of light, $k$ denotes the curvature taking one of three values $\{-1,~0,~1\}$. The symbol $a(t)$ is the dimensionless scale factor normalized to unity at the present time $(a(t_0) = 1)$.  \\

To investigate the expansion history in a model-independent cosmographic framework, we adopt the approach of reference \cite{1972gcpa.book.....W}. The expression of the scale factor as a Taylor expansion around the present cosmic time is given as 

\begin{equation}\label{equ_expansion}
\frac{a(t)}{a\left(t_0\right)} =1+H_0\left(t-t_0\right)-\frac{q_0}{2} H_0^2\left(t-t_0\right)^2
 +\frac{j_0}{3!} H_0^3\left(t-t_0\right)^3+\frac{s_0}{4!} H_0^4\left(t-t_0\right)^4+\mathcal{O}\left[\left(t-t_0\right)\right]^5
\end{equation}
where, the cosmographic parameters, $H_0$,  shown in Equation \ref{equ_expansion} are expressed as

\begin{equation}\label{equ_cosmog_param}
H_0=\left.\frac{\dot{a}}{a}\right|_{t=t_0}, \qquad
q_0=-\left.\frac{1}{H_0^2}\,\frac{\ddot{a}}{a}\right|_{t=t_0}, \qquad
j_0=\left.\frac{1}{H_0^3}\,\frac{\dot{\ddot{a}}}{a}\right|_{t=t_0}, \qquad
s_0=\left.\frac{1}{H_0^4}\,\frac{\ddot{\ddot{a}}}{a}\right|_{t=t_0}.
\end{equation}

Thus, the Taylor-expansion of Hubble parameter is    

\begin{equation}\label{equ_h_z_series}
    H(z) = H_0 + \left.\frac{dH}{dz}\right|_{z=0} z 
           + \left.\frac{d^2H}{dz^2}\right|_{z=0} \dfrac{z^2}{2!} 
           + \left.\frac{d^3H}{dz^3}\right|_{z=0} \dfrac{z^3}{3!}  
           + \left.\frac{d^4H}{dz^4}\right|_{z=0} \dfrac{z^4}{4!} 
\end{equation}

The Taylor series expansion in terms of the standard redshift $z$ remains valid only for the low-redshift regime ($z < 1$), while a significant portion of the recent observational data lies in the high-redshift domain ($z > 1$). In fact, the radius of convergence of any expansion in powers of $z$ is at most of order unity, which causes the series to diverge or lose accuracy when $z > 1$. To address this limitation, an alternative redshift variable, the $y$-redshift, defined as $y = \frac{z}{1+z}$, has been introduced in the literature \cite{2011PhRvD..84l4061C}. Although this change of variable does not alter the underlying physics, it substantially improves the convergence properties of the expansion. In the $y$-redshift parametrization, the radius of convergence extends to $|y| < 1$, corresponding to the full range $0 \leq z < \infty$. Consequently, the Taylor expansion of the Hubble parameter in terms of $y$ can be reliably applied to both low- and high-redshift regimes, and is expressed as follows \cite{2011PhRvD..84l4061C}:

\begin{equation}\label{equ_h_y_series}
    H(y) = H_0 + \left.\frac{dH}{dy}\right|_{y=0} y
           + \left.\frac{d^2H}{dy^2}\right|_{y=0} y^2
           + \left.\frac{d^3H}{dy^3}\right|_{y=0} y^3
           + \left.\frac{d^4H}{dy^4}\right|_{y=0} y^4
\end{equation}

Using Equations \ref{equ_expansion} and \ref{equ_h_y_series}, one can express the Hubble parameter, and cosmological distances in terms of Cosmographic parameters $(q_0,~j_0,~s_0)$ as 

\begin{itemize}
    \item  Hubble Parameter: $H_0(y)=H_0E(y)$
        \begin{equation}
            E(y) = 1 + (1 + q_0)y + \frac{1}{2}(-q_0^2 + 2q_0 + j_0 + 2)y^2 + \frac{1}{6}(6 + 3q_0^3 - 3q_0^2 + 6q_0 + 3j_0 - 4j_0q_0 - s_0)y^3
        \end{equation}
        
    \item Hubble Distance: $d_H = \dfrac{c}{H_0}D_H$
        \begin{equation}
            D_\mathrm{H}(y) = 1 - (1 + q_0)y + \frac{1}{2}(3q_0^2 + 2q_0 - j_0)y^2 + \frac{1}{6}(3j_0 + 10j_0q_0 - 9q_0^2 - 15q_0^3 + s_0)y^3 
        \end{equation}
        
    \item Transverse Comoving Distance: $d_{\mathrm{co}}=\dfrac{c}{H_0}D_{\mathrm{co}}=\frac{c}{H_0\sqrt{|\Omega_{k0}|}}{\sin}{n}\left(\sqrt{|\Omega_{k0}|}\displaystyle\int_1^y \frac{D_{\mathrm{H}}}{\left(1-y\right)^2}dy\right)$

        \begin{align}
            D_{\mathrm{co}}(y) &= y + \frac{1}{2}(1 - q_0)y^2 + \frac{1}{6}(2 - 2q_0 + 3q_0^2 - j_0+\Omega_{k0})y^3 \nonumber \\
                   &\quad + \frac{1}{24}(6 - 6q_0 + 9q_0^2 - 15q_0^3 + j_0(-3 + 10q_0) + s_0+6\Omega_{k0}-6\Omega_{k0}q_0)y^4
        \end{align}

    \item Angular Diameter Distance:     $d_{\mathrm{A}}=\dfrac{c}{H_0}D_{\mathrm{A}}$
    \begin{equation}
          D_\mathrm{A}=D_\mathrm{co}\times\left(1-y\right)
    \end{equation}

    \item Luminosity Distance:     $d_{\mathrm{L}}=\dfrac{c}{H_0}D_{\mathrm{L}}$
    \begin{equation}
          D_\mathrm{L}=\dfrac{D_\mathrm{co}}{\left(1-y\right)}
    \end{equation}

    \item Distance Modulus:  
    \begin{equation}
          \mu=5\log\left[\dfrac{c/H_0\times D_L}{1~\mathrm{Mpc}}\right]+25
    \end{equation}
    
    \item Volume Distance: $d_V=\dfrac{c}{H_0}D_V=\dfrac{c}{H_0}\left(zD_HD^2_{\mathrm{co}}\right)^{1/3}$

    \begin{align}
        D_V &= y + \frac{1}{3} \left(1 - 2q_0\right) y^2 + \frac{1}{36} \left(7 - 10j_0 + 4\Omega_{k0} - 10q_0 + 29q_0^2\right) y^3 \nonumber \\
        &\quad + \frac{1}{324} \bigl(44 + \Omega_{k0}(48 - 60q_0) - 57q_0 + 117q_0^2 - 376q_0^3 + 3j_0(-13 + 86q_0) + 27s_0\bigr) y^4 
    \end{align}
\end{itemize}

\subsection{Distance Sum Rule (DSR)}
Under the assumptions of homogeneity and isotropy, the dimensionless comoving distances $(D_\mathrm{co})$ are defined as  
\begin{equation}\label{eq:sl7b}
\begin{aligned}
&D_\mathrm{co}^{^\mathrm{os}} \equiv D_\mathrm{co}(0,z_s) \equiv \dfrac{H_0}{c}d_\mathrm{co}^{^\mathrm{os}}, \\
&D_\mathrm{co}^{^\mathrm{ol}} \equiv D_\mathrm{co}(0,z_l) \equiv \dfrac{H_0}{c}d_\mathrm{co}^{^\mathrm{ol}}, \\
&D_\mathrm{co}^{^\mathrm{ls}} \equiv D_\mathrm{co}(z_l,z_s) \equiv \dfrac{H_0}{c}d_\mathrm{co}^{^\mathrm{ls}},
\end{aligned}
\end{equation}

In a spatially flat universe, the dimensionless comoving distances satisfy a simple additivity relation, namely $D_\mathrm{co}^{^\mathrm{os}} = D_\mathrm{co}^{^\mathrm{ol}} + D_\mathrm{co}^{^\mathrm{ls}}$.  For non-flat geometries, however, these distances are related  through the generalized Distance Sum Rule (DSR) \cite{1993ppc..book.....P,2015PhRvL.115j1301R}, which incorporates the effect of spatial curvature. This relation can be expressed as

\begin{equation}\label{equ_dsr}
D_\mathrm{co}^{^{\mathrm{ls}}} = D_\mathrm{co}^{^{\mathrm{os}}} \sqrt{1 + \Omega_{k0} \, \left(D_\mathrm{co}^{^{\mathrm{ol}}}\right)^2} - D_\mathrm{co}^{^{\mathrm{ol}}} \sqrt{1 + \Omega_{k0} \, \left(D_\mathrm{co}^{^{\mathrm{os}}}\right)^2}
\end{equation}

Further, we can write the distance sum rule in terms of the time-delay distance as
\begin{equation}\label{eq:sl9}
\begin{aligned}
d_{\Delta t}^{\mathrm{th}}=&\left(1+z_l\right)\frac{{d_\mathrm{A}^{^\mathrm{o s}} d_\mathrm{A}^{^\mathrm{o l}}}}{{d_\mathrm{A}^{^\mathrm{l s}}}}\\=&\dfrac{c}{H_0}\dfrac{D_{\mathrm{co}}^{^{\mathrm{ol}}}D_{\mathrm{co}}^{^{\mathrm{os}}}}{D_{\mathrm{co}}^{^{\mathrm{ls}}}} =\dfrac{c}{H_0}
\left[ \dfrac{1}{D_{\mathrm{co}}^{^{\mathrm{ol}}}} \sqrt{1+\Omega_{k0}\left(D_{\mathrm{co}}^{^{\mathrm{ol}}}\right)^{2}}
- \dfrac{1}{D_{\mathrm{co}}^{^{\mathrm{os}}}} \sqrt{1+\Omega_{k0}\left(D_{\mathrm{co}}^{^{\mathrm{os}}}\right)^{2}} \right]^{-1}.
\end{aligned}
\end{equation}     

The spatial curvature parameter $\Omega_{k0}$ can therefore be determined directly from Equation \ref{eq:sl9} without adopting a fiducial cosmological model, provided that $D_{\mathrm{co}}^{^{\mathrm{ol}}}$ and $D_{\mathrm{co}}^{^{\mathrm{os}}}$ are obtained from cosmography or other observations. This  equation represents the theoretical formulation of the time-delay distance. To evaluate the left-hand side of these relations, we use time-delay distance data of strong gravitational lensing.  

\subsection{Statistical Analysis}

In this work, we perform a Bayesian analysis to constrain cosmological parameters by comparing theoretical predictions with observational data. The analysis involves computing chi-square functions for each dataset separately and then combining them to derive the overall likelihood. The datasets considered include time-delay distance measurements from strong gravitational lensing systems (SGL), Type Ia Supernovae (SNIa), and Baryon Acoustic Oscillations (BAO). Below, we describe the chi-square expressions for each observational dataset in detail.

\subsubsection*{Time-delay Distance from Strong Gravitational Lensing}

{The chi-square for the time-delay distance for 5 H0LiCOW and DES J0408$-$5354 samples in SGL systems is defined as:}

\begin{equation}\label{equ_chi_sgl}
    \chi^2_{\mathrm{TDD}}\left(H_0,\Omega_{k0},q_0,j_0,s_0\right)=\sum_i\left[\dfrac{d_{\Delta t}^{\mathrm{th}}\left(z_i;H_0,\Omega_{k0},q_0,j_0,s_0\right)-d_{\Delta t}^{\mathrm{obs}}\left(z_i\right)}{\sigma_{d_{\Delta t}^{\mathrm{obs}}}\left(z_i\right)}\right]^2,
\end{equation}

where $d_{\Delta t}^{\mathrm{th}}$ denotes the theoretically computed time-delay distance as defined in Equation \ref{eq:sl9}. The observed time-delay distance and its associated uncertainty are represented by $d_{\Delta t}^{\mathrm{obs}}$ and $\sigma_{d_{\Delta t}^{\mathrm{obs}}}$, respectively.\\

Finally, we define the total chi-square for the strong gravitational lensing analysis as
\begin{equation}
    \chi^2_{\mathrm{SGL}} = \chi^2_{\mathrm{TDD}} + \chi^2_{\mathrm{B1608+656}} + \chi^2_{\mathrm{SDSSJ0946+1006}},
\end{equation}
where $\chi^2_{\mathrm{TDD}}$ is given in Equation \ref{equ_chi_sgl}, $\chi^2_{\mathrm{B1608+656}}$ is obtained from the likelihood in Equation \ref{equ_likel_b1608}, and $\chi^2_{\mathrm{SDSSJ0946+1006}}$ is obtained from the likelihood in Equation \ref{equ_likeli_beta}.

\subsubsection*{Type Ia Supernovae}

The theoretical apparent magnitude for SNIa is given by: 

\begin{align}\label{equ_m_th}
    m^{\mathrm{th}} &= 5\log_{10}\left(\dfrac{d_L}{1\,\mathrm{Mpc}}\right) + 25 + M_B \nonumber\\
                    &= 5\log_{10}\left(\dfrac{c}{H_0}D_L\right) + 25 + M_B \nonumber\\
                    &= 5\log_{10}\left(D_L\right) + \mathcal{M},
\end{align}
where $\mathcal{M}=5\log_{10}\left(\dfrac{c}{H_0}\right)+M_B+25$. In this analysis, we treat $\mathcal{M}$ as a nuisance parameter and numerically marginalize over it using a wide prior distribution.\\    

\noindent After marginalizing over $\mathcal{M}$, the chi-square for SNIa data is:

\begin{equation}
    \chi^2_{\mathrm{SNIa}}\left(\Omega_{k0},q_0,j_0,s_0\right) = a - \dfrac{b^2}{f} + \ln\left(\dfrac{f}{2\pi}\right),
\end{equation}

where:

\begin{align}
    a &= \Delta\bar{m}^T \cdot \mathrm{Cov}^{-1} \cdot \Delta\bar{m}, \nonumber\\
    b &= \Delta\bar{m}^T \cdot \mathrm{Cov}^{-1} \cdot I, \nonumber\\
    f &= I \cdot \mathrm{Cov}^{-1} \cdot I, \nonumber\\
    \Delta\bar{m} &= m^{\mathrm{obs}} - 5\log_{10}\left(D_L\right).
\end{align}
{In this expression the terms $\mathrm{Cov}$ denotes the covariance matrix and $I$ indicates a vector of ones (or the identity matrix in the covariance formalism).}
\subsubsection*{Baryon Acoustic Oscillations}

The chi-square function for the BAO dataset takes the form:   

\begin{equation}
\begin{split}
    \chi_{\mathrm{BAO}}^2\left(H_0,\Omega_{k0},q_0,j_0,s_0\right) = &\left[{X_{\mathrm{th}}\left(z;H_0,\Omega_{k0},q_0,j_0,s_0\right)-X_{\mathrm{obs}}\left(z\right)}\right]^T \operatorname{Cov}^{-1} \\
    &\left[{X_{\mathrm{th}}\left(z;H_0,\Omega_{k0},q_0,j_0,s_0\right)-X_{\mathrm{obs}}\left(z\right)}\right],
\end{split}
\end{equation}
where the vector $X$ defines the relevant BAO measurements defined in Equations \ref{eq_dM}, \ref{eq_dH}, and \ref{eq_dV}.
\subsubsection*{Total Chi-square and Likelihood}

The total chi-square function combines the contributions from all independent datasets as     

\begin{equation}
    \chi^2_T = \chi^2_{\mathrm{SGL}} + \chi^2_{\mathrm{SNIa}} + \chi^2_{\mathrm{BAO}},
\end{equation}

where $\chi^2_{\mathrm{SNIa}}$ corresponds to the specific supernova samples considered in the analysis, namely $\chi^2_{\mathrm{PantheonPlus}}$, $\chi^2_{\mathrm{Union3}}$, and $\chi^2_{\mathrm{DESY5}}$. \\ 

The corresponding likelihood function is given by

\begin{equation}
    \mathcal{L} \propto \exp \left(-\frac{\chi_T^2}{2} \right).
\end{equation}

Since the datasets are independent, the individual chi-square terms are additive. This framework allows for a joint analysis that provides statistically robust constraints on the cosmological parameters.

\subsubsection*{MCMC Implementation and Priors}
We adopt flat priors for all cosmological parameters, as summarized in Table \ref{tab_prior}. The Markov Chain Monte Carlo (MCMC) analysis uses 12 walkers with 10,000 steps each to thoroughly explore the parameter space. The first 30\% of samples are discarded as burn-in, and the posterior distributions are analysed using the remaining samples. To ensure convergence, we calculate the integrated auto-correlation time $\tau_f$ using the function \textbf{\textit{autocorr.integrated\_time}} from the \textbf{\textit{emcee}} package. For further details on this procedure, please refer to reference \cite{2013PASP..125..306F}.

\begin{table}[h]
    \centering
    \renewcommand{\arraystretch}{2}
    \begin{tabular}[b]{|l|l|}
        \hline
        Parameter & Prior Range \\
        \hline \hline
        $H_0~[\mathrm{km\,s^{-1}\,Mpc^{-1}}]$ & $\mathbb{U}[0,100]$ \\
        \hline
        $\Omega_{k0}$ & $\mathbb{U}[-0.5,1]$ \\
        \hline
        $q_0$ & $\mathbb{U}[-2,2]$ \\
        \hline
        $j_0$ & $\mathbb{U}[-3,3]$ \\
        \hline
        $s_0$ & $\mathbb{U}[-10,10]$ \\
        \hline
    \end{tabular}
    \caption{The prior range for $H_0$, $\Omega_{k0}$, and cosmographic parameters.}
    \label{tab_prior}
\end{table}

\section{Results}\label{sec_resu}   

In this section, we present the constraints on the cosmographic and cosmological parameters derived from various observational datasets. The analysis is divided into two subsections: results obtained without including the DESI-DR2 dataset and those with its inclusion. For each case, we consider combinations of strong gravitational lensing with different Type Ia supernova datasets, namely PantheonPlus, Union3, and DESY5. The parameter constraints for $H_0$, $\Omega_{k0}$, $q_0$, $j_0$, and $s_0$ are shown along with their corresponding $68\%$ and $95\%$ confidence intervals. The central values and trends are discussed with reference to the standard $\Lambda$CDM expectations, where $q_0 \approx -0.55$, $j_0 = 1$, $s_0 \approx -0.35$ and $\Omega_{k0} = 0$.

\subsection{Results Without DESI-DR2}   

The best fit value of parameters obtained from the combination of SGL with different SNIa datasets excluding DESI-DR2 are tabulated in Table \ref{tab:cosmography_results_without_desi}. For the SGL+PantheonPlus combination, we find $H_0 = 75.700^{+1.957}_{-1.806}$ km\,s$^{-1}$\,Mpc$^{-1}$, which is consistent with the H0LiCOW survey and the value obtained from SH0ES \cite{2022ApJ...934L...7R} at  68\% confidence level. The curvature parameter $\Omega_{k0} = 0.165^{+0.116}_{-0.102}$ suggests a mild preference for an open universe, although it is consistent with a flat universe within 95\% confidence level. The deceleration parameter $q_0 = -0.453^{+0.070}_{-0.066}$ is slightly less negative than the $\Lambda$CDM expectation, which points to a marginally weaker cosmic acceleration. The jerk parameter $j_0 = 0.810^{+0.732}_{-0.741}$ is consistent with $j_0 = 1$ at 68\% confidence level, while the snap parameter $s_0 = -0.664^{+7.044}_{-6.462}$ is poorly constrained. \\ 

For the SGL+Union3 dataset, the Hubble constant is constrained as $H_0 = 74.912^{+2.015}_{-1.805}$ km\,s$^{-1}$\,Mpc$^{-1}$, higher than the H0LiCOW survey value but consistent within 68\% confidence level. The curvature parameter $\Omega_{k0} = 0.253^{+0.132}_{-0.116}$ suggests a stronger inclination towards open universe, though compatible with flatness at 95\% confidence level. The deceleration parameter $q_0 = -0.219^{+0.140}_{-0.122}$ is significantly less negative than expected, which indicates a slower expansion rate. The jerk parameter $j_0 = -0.383^{+1.035}_{-1.119}$ is broadly consistent with the standard model but shows larger uncertainty. The snap parameter $s_0 = -2.778^{+7.446}_{-5.137}$ remains weakly constrained. \\

For the SGL+DESY5 dataset, the constraints are similar to those from the PantheonPlus combination. We find $H_0 = 75.642^{+1.969}_{-1.829}$ km\,s$^{-1}$\,Mpc$^{-1}$, with $\Omega_{k0} = 0.157^{+0.107}_{-0.098}$ and $q_0 = -0.456^{+0.071}_{-0.068}$, indicating a preference for an open universe and a deceleration parameter close to that of $\Lambda$CDM. The jerk parameter $j_0 = 0.839^{+0.702}_{-0.724}$ is consistent with unity, while the snap parameter $s_0 = -0.337^{+6.878}_{-6.556}$ is again poorly constrained.

\begin{table}[htbp]         
\renewcommand{\arraystretch}{2}
\centering  
\begin{tabular}{|l|c|c|c|c|c|}
\hline
Dataset & $H_0~[\mathrm{km\,s^{-1}\,Mpc^{-1}}]$ & $\Omega_{k0}$ & $q_0$ & $j_0$ & $s_0$\\
\hline
SGL+PantheonPlus & $75.700^{+1.957}_{-1.806}$ & $0.165^{+0.116}_{-0.102}$ & $-0.453^{+0.070}_{-0.066}$ & $0.810^{+0.732}_{-0.741}$ & $-0.664^{+7.044}_{-6.462}$\\
SGL+Union3 & $74.912^{+2.015}_{-1.805}$ & $0.253^{+0.132}_{-0.116}$ & $-0.219^{+0.140}_{-0.122}$ & $-0.383^{+1.035}_{-1.119}$ & $-2.778^{+7.446}_{-5.137}$\\
SGL+DESY5 & $75.642^{+1.969}_{-1.829}$ & $0.157^{+0.107}_{-0.098}$ & $-0.456^{+0.071}_{-0.068}$ & $0.839^{+0.702}_{-0.724}$ & $-0.337^{+6.878}_{-6.556}$\\
\hline
\end{tabular}
\caption{Constraints on cosmological and cosmographic parameters from SGL combined with various SNIa datasets.}
\label{tab:cosmography_results_without_desi}
\end{table}

\subsection{Results With DESI-DR2}

When DESI-DR2 data is included, the constraints on parameters tighten considerably as shown in Table \ref{tab:cosmography_results_with_desi}. For SGL+PantheonPlus+DESI-DR2, the Hubble constant is tightly constrained as $H_0 = 68.398^{+0.433}_{-0.444}$ km\,s$^{-1}$\,Mpc$^{-1}$, which is closer to the Planck $\Lambda$CDM value. The curvature parameter $\Omega_{k0} = -0.036^{+0.045}_{-0.046}$ is consistent with flatness at 68\% confidence level. The deceleration parameter $q_0 = -0.477^{+0.054}_{-0.058}$ aligns well with $\Lambda$CDM expectations. The jerk parameter $j_0 = 0.835^{+0.427}_{-0.394}$ is also consistent with unity, while the snap parameter $s_0 = 0.629^{+2.249}_{-1.786}$ is constrained better than in the previous case.\\

For the SGL+Union3+DESI-DR2 combination, we obtain $H_0 = 67.461^{+0.572}_{-0.645}$ km\,s$^{-1}$\,Mpc$^{-1}$ and $\Omega_{k0} = -0.033^{+0.045}_{-0.047}$, again consistent with flat geometry. The deceleration parameter $q_0 = -0.350^{+0.077}_{-0.077}$ is closer to $\Lambda$CDM than the non-DESI case but slightly less negative. The jerk parameter $j_0 = 0.079^{+0.482}_{-0.453}$ includes unity within 95\% confidence level, while the snap parameter $s_0 = -2.262^{+1.775}_{-1.128}$ shows improved constraints.\\

For the SGL+DESY5+DESI-DR2 dataset, we find $H_0 = 68.720^{+0.520}_{-0.633}$ km\,s$^{-1}$\,Mpc$^{-1}$ and $\Omega_{k0} = -0.054^{+0.050}_{-0.046}$, again consistent with flatness. The deceleration parameter $q_0 = -0.546^{+0.087}_{-0.081}$ is in excellent agreement with $\Lambda$CDM expectations. The jerk parameter $j_0 = 1.344^{+0.628}_{-0.616}$ overlaps with unity within uncertainties, while the snap parameter $s_0 = 3.335^{+3.951}_{-3.191}$, despite its larger uncertainty, provides a meaningful constraint.

\begin{table}[htbp]
\renewcommand{\arraystretch}{2}
\centering
\begin{tabular}{|l|c|c|c|c|c|}
\hline
Dataset & $H_0~[\mathrm{km\,s^{-1}\,Mpc^{-1}}]$ & $\Omega_{k0}$ & $q_0$ & $j_0$ & $s_0$\\
\hline
\makecell[l]{SGL+PantheonPlus\\~~~~~~+DESI-DR2} & $68.398^{+0.433}_{-0.444}$ & $-0.036^{+0.045}_{-0.046}$ & $-0.477^{+0.054}_{-0.058}$ & $0.835^{+0.427}_{-0.394}$ & $0.629^{+2.249}_{-1.786}$\\
\makecell[l]{SGL+Union3\\~~~~~~+DESI-DR2} & $67.461^{+0.572}_{-0.645}$ & $-0.033^{+0.045}_{-0.047}$ & $-0.350^{+0.077}_{-0.077}$ & $0.079^{+0.482}_{-0.453}$ & $-2.262^{+1.775}_{-1.128}$\\
\makecell[l]{SGL+DESY5\\~~~~~~+DESI-DR2} & $68.720^{+0.520}_{-0.633}$ & $-0.054^{+0.050}_{-0.046}$ & $-0.546^{+0.087}_{-0.081}$ & $1.344^{+0.628}_{-0.616}$ & $3.335^{+3.951}_{-3.191}$\\
\hline
\end{tabular}
\caption{Constraints on cosmological and cosmographic parameters from SGL combined with various SNIa datasets including DESI-DR2.}
\label{tab:cosmography_results_with_desi}
\end{table}

\subsection{Comparison of Contours and Parameter Degeneracies}


\begin{figure}[htbp]
    \centering

    \subfloat[Confidence contours and posterior distributions from SGL+PantheonPlus and SGL+PantheonPlus+DESI-DR2 datasets.]
    {\includegraphics[width=0.98\linewidth]{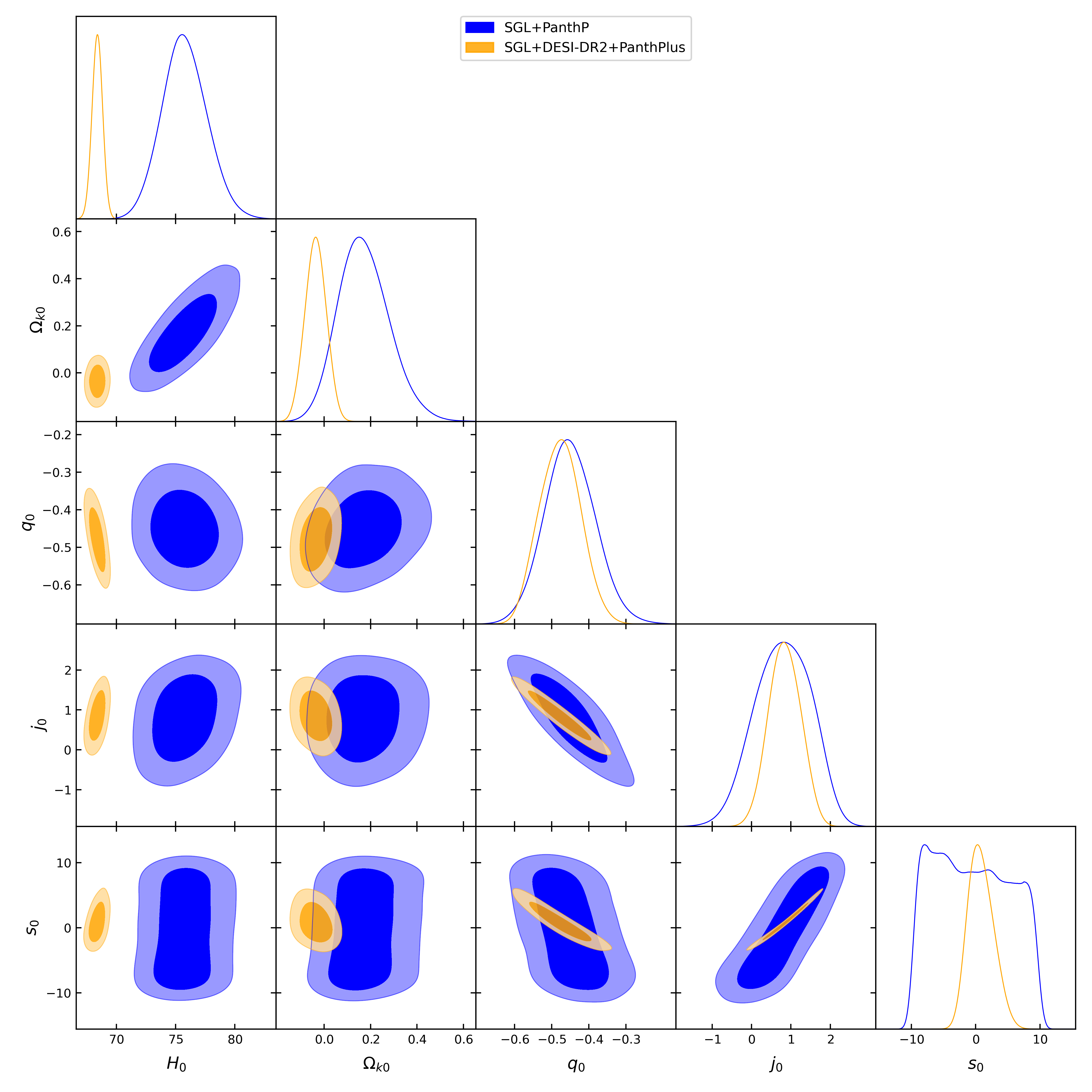}
    \label{fig_cntr_panthp_desi}}\\[6pt]

    \subfloat[Correlation matrix: SGL+PantheonPlus]
    {\includegraphics[width=0.45\linewidth]{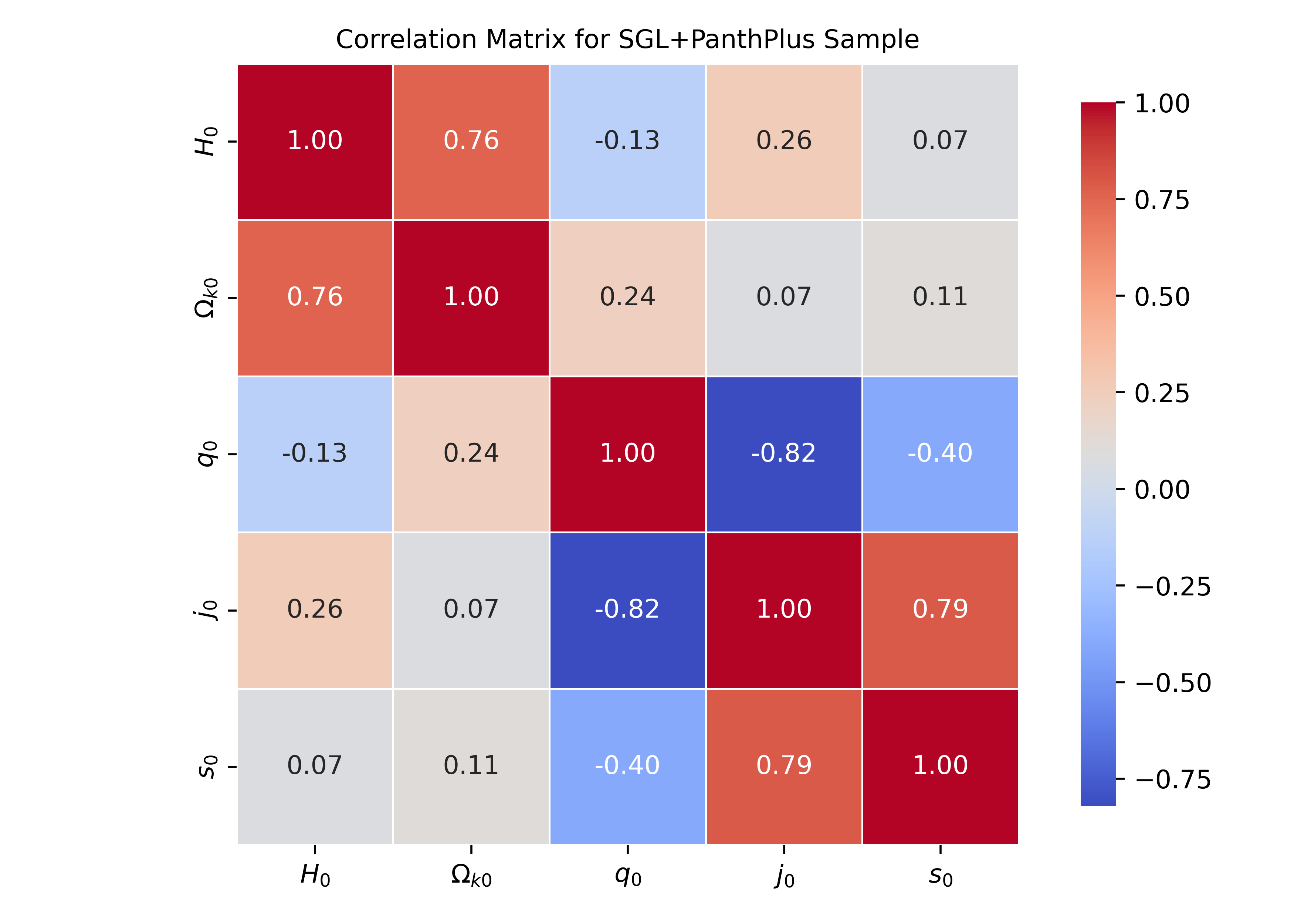}
    \label{fig_corr_panthp}}
    \hspace{0.05\linewidth}
    \subfloat[Correlation matrix: SGL+ PantheonPlus+DESI-DR2]
    {\includegraphics[width=0.45\linewidth]{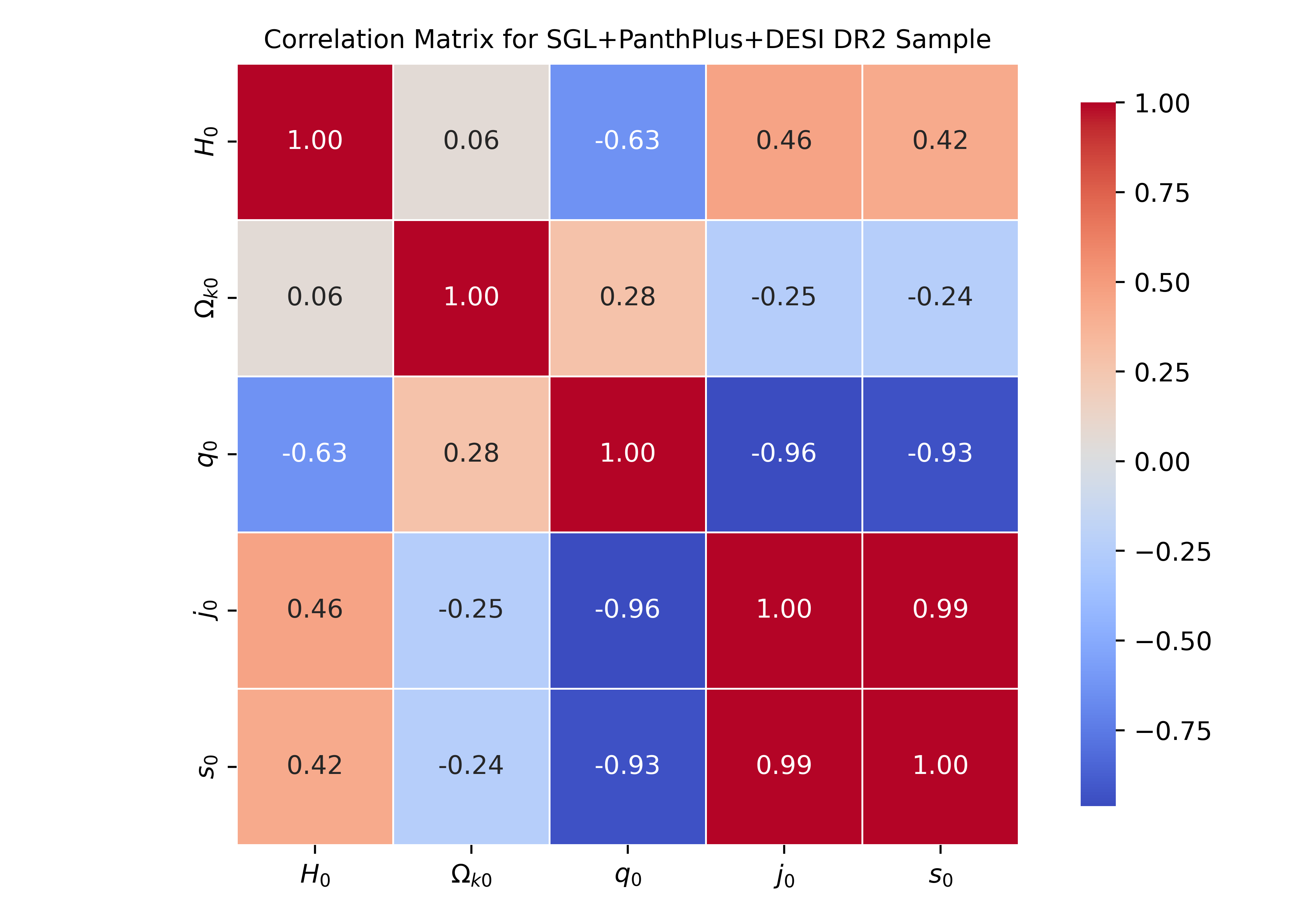}
    \label{fig_corr_panthp_desi}}

       \caption{(a) Posterior contours for SGL+PantheonPlus and SGL+PantheonPlus+DESI-DR2 datasets; (b)–(c) corresponding correlation matrices for each dataset combination.}

    \label{fig_combined_contour_corr_panthp}
\end{figure}


\begin{figure}[htbp]
    \centering

    \subfloat[Confidence contours and posterior distributions from SGL+Union3 and SGL+Union3+DESI-DR2 datasets.]
    {\includegraphics[width=0.98\linewidth]{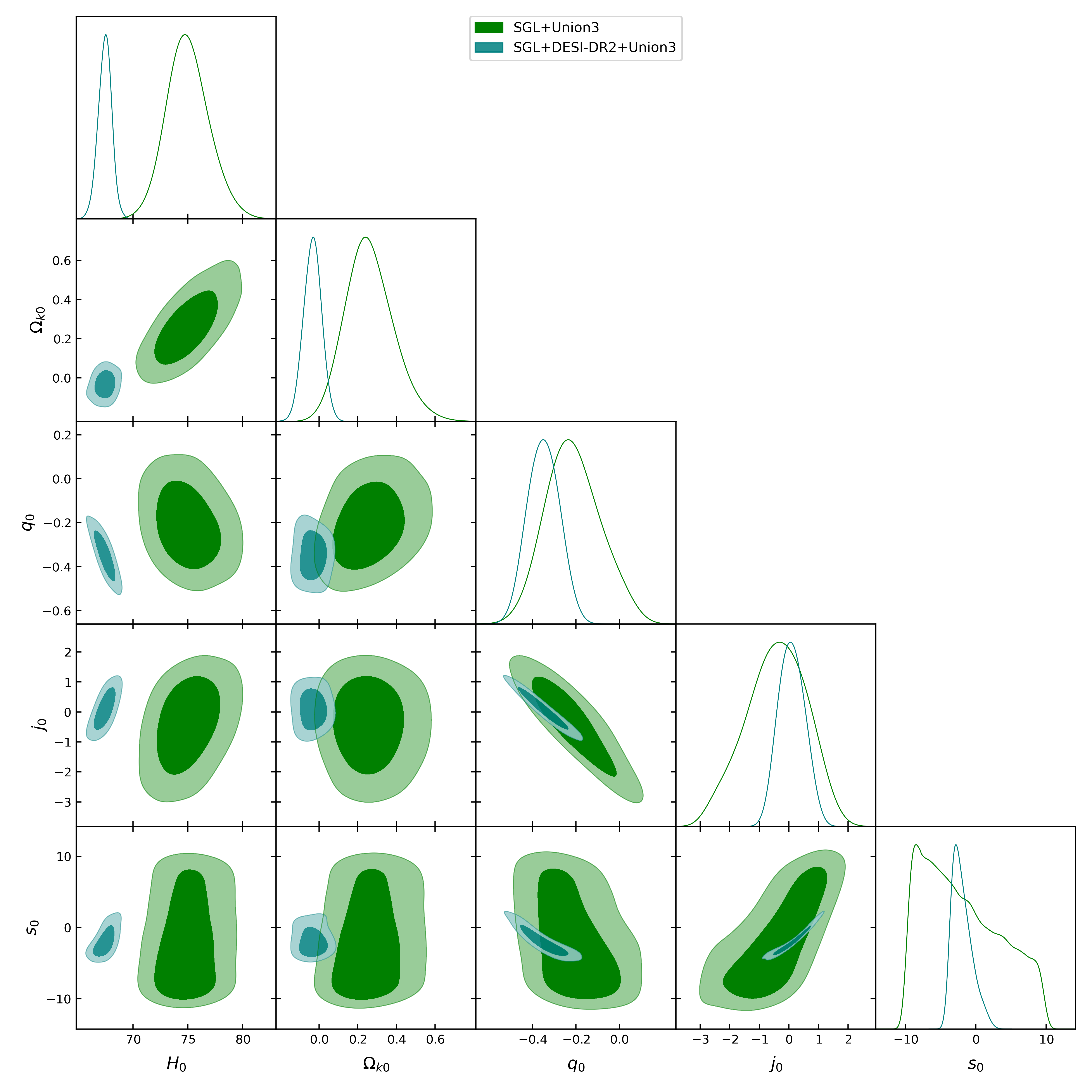}
    \label{fig_cntr_union3_desi}}\\[6pt]

    \subfloat[Correlation matrix: SGL+Union3]
    {\includegraphics[width=0.45\linewidth]{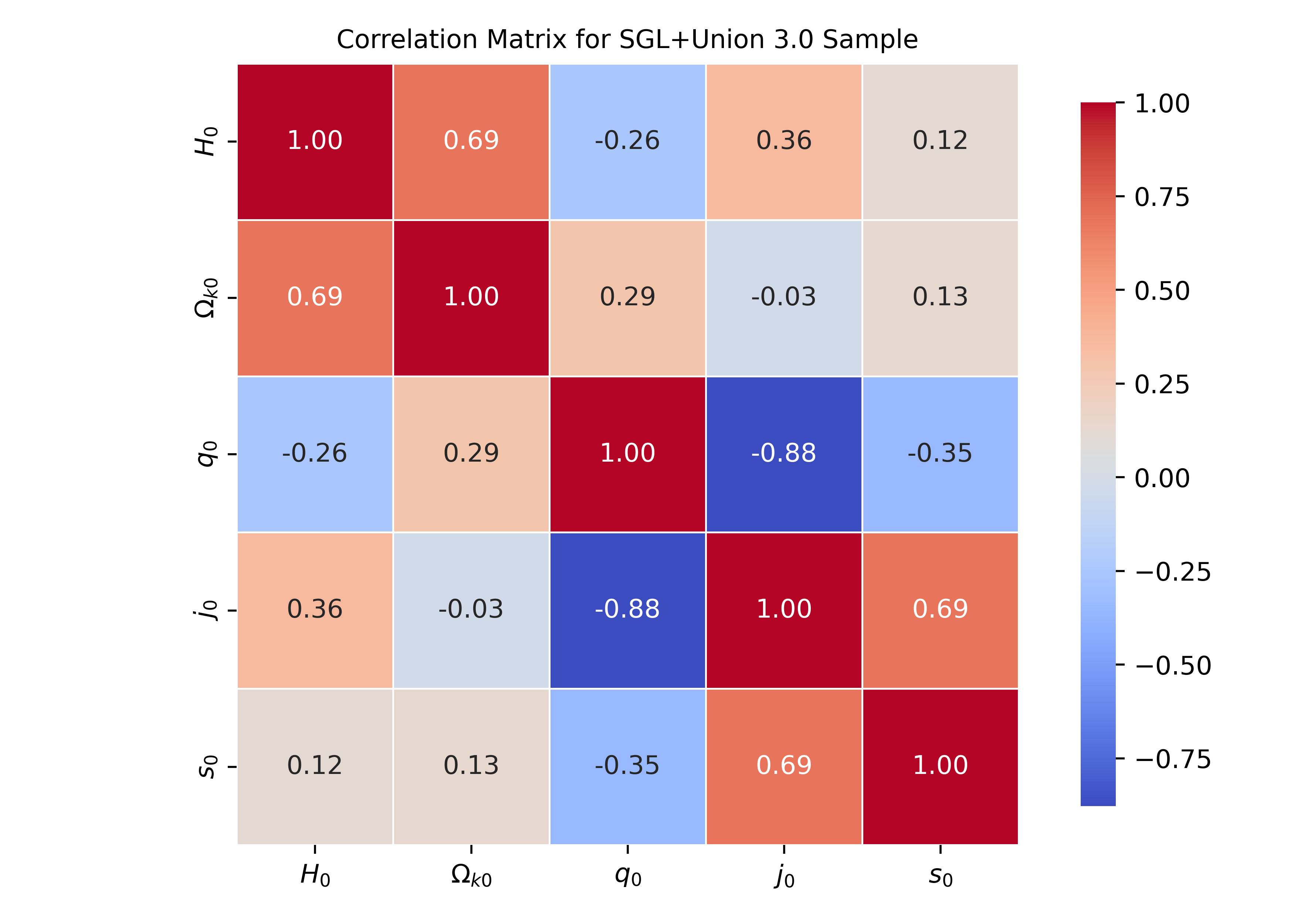}
    \label{fig_corr_union3}}
    \hspace{0.05\linewidth}
    \subfloat[Correlation matrix: SGL+ Union3+DESI-DR2]
    {\includegraphics[width=0.45\linewidth]{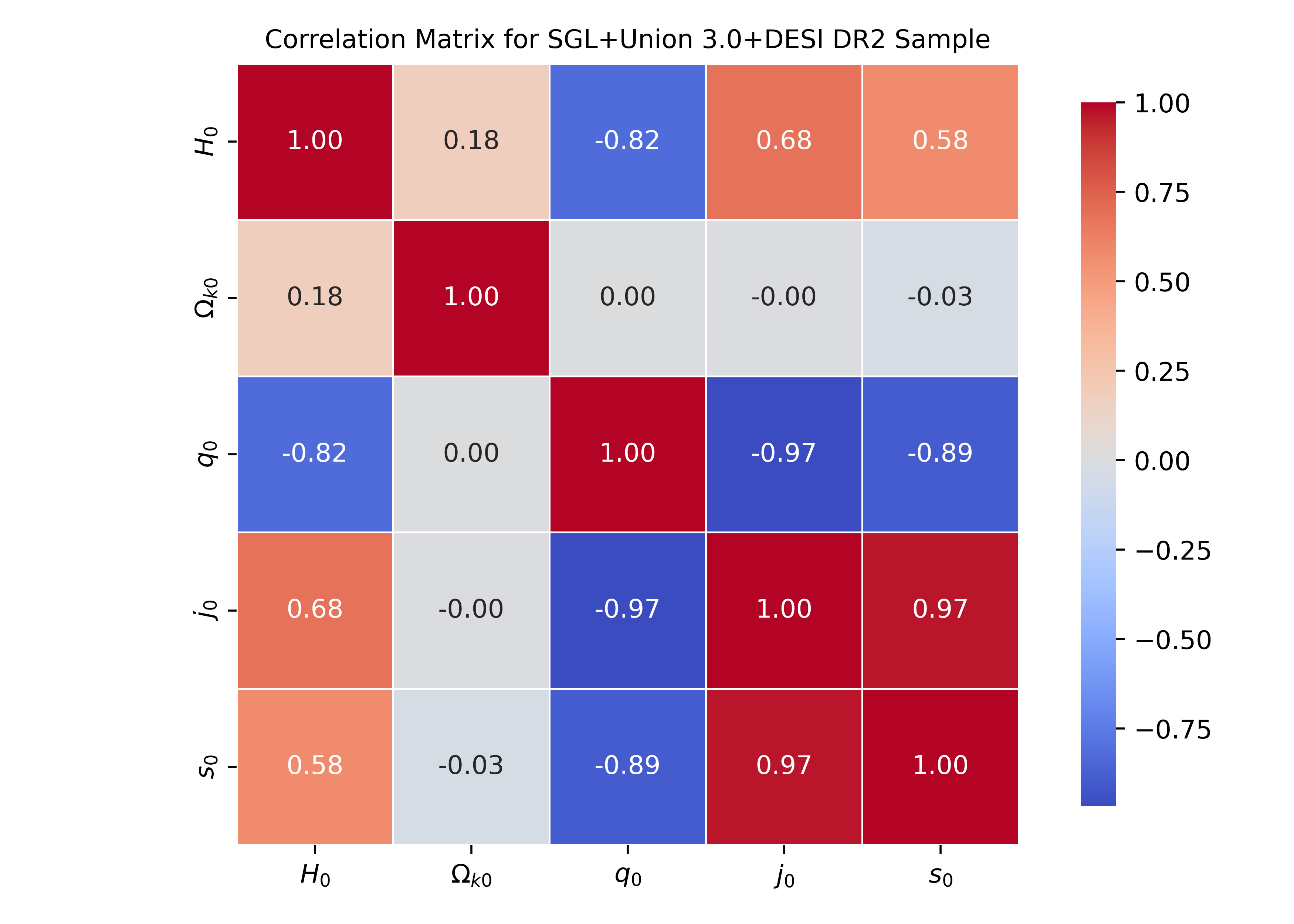}
    \label{fig_corr_union3_desi}}

       \caption{(a) Posterior contours for SGL+Union3 and SGL+Union3+DESI-DR2 datasets; (b)–(c) corresponding correlation matrices for each dataset combination.}

    \label{fig_combined_contour_corr_union3}
\end{figure}


\begin{figure}[htbp]
    \centering

    \subfloat[Confidence contours and posterior distributions from SGL+DESY5 and SGL+DESY5+DESI-DR2 datasets.]
    {\includegraphics[width=0.98\linewidth]{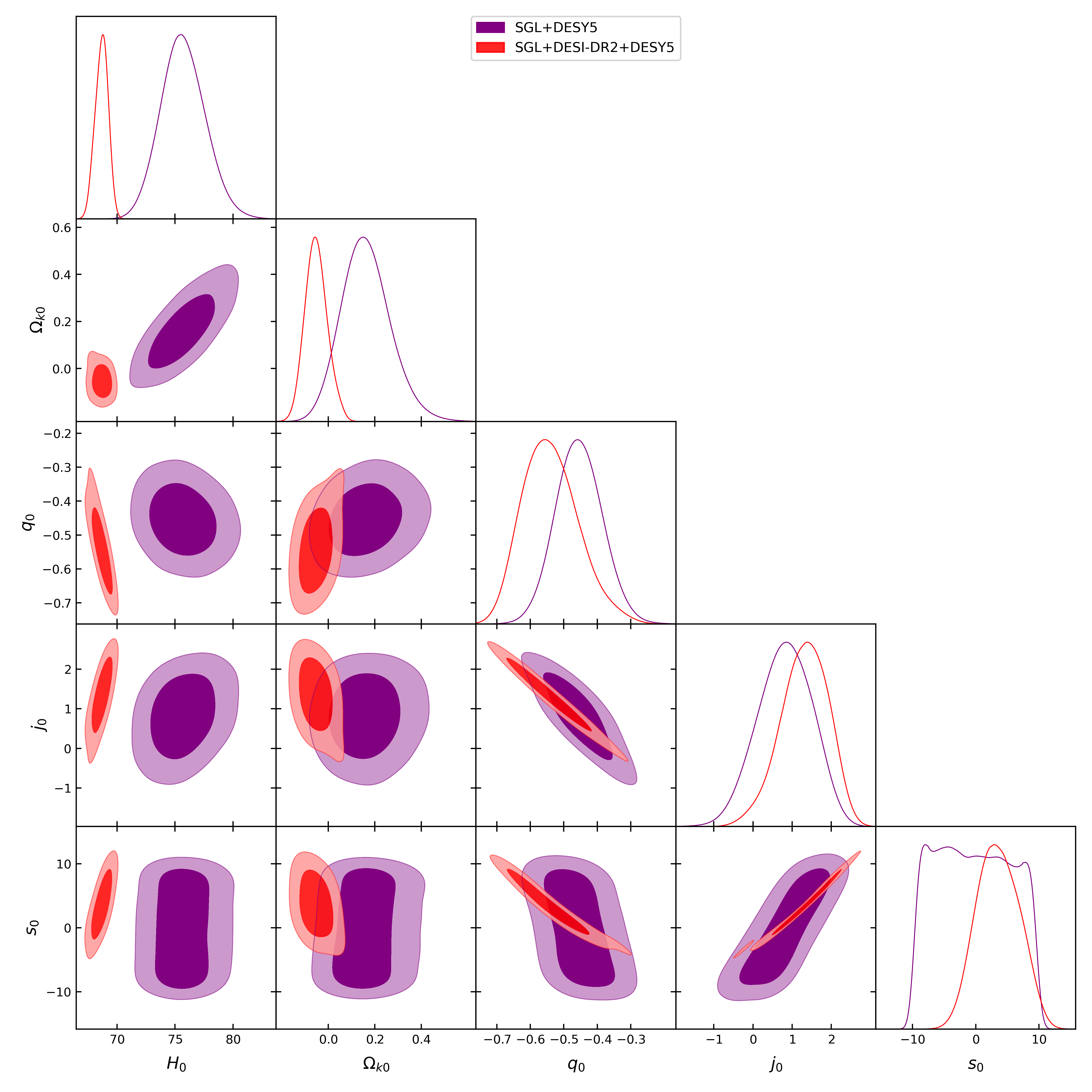}
    \label{fig_cntr_desy5_desi}}\\[6pt]

    \subfloat[Correlation matrix: SGL+DESY5]
    {\includegraphics[width=0.45\linewidth]{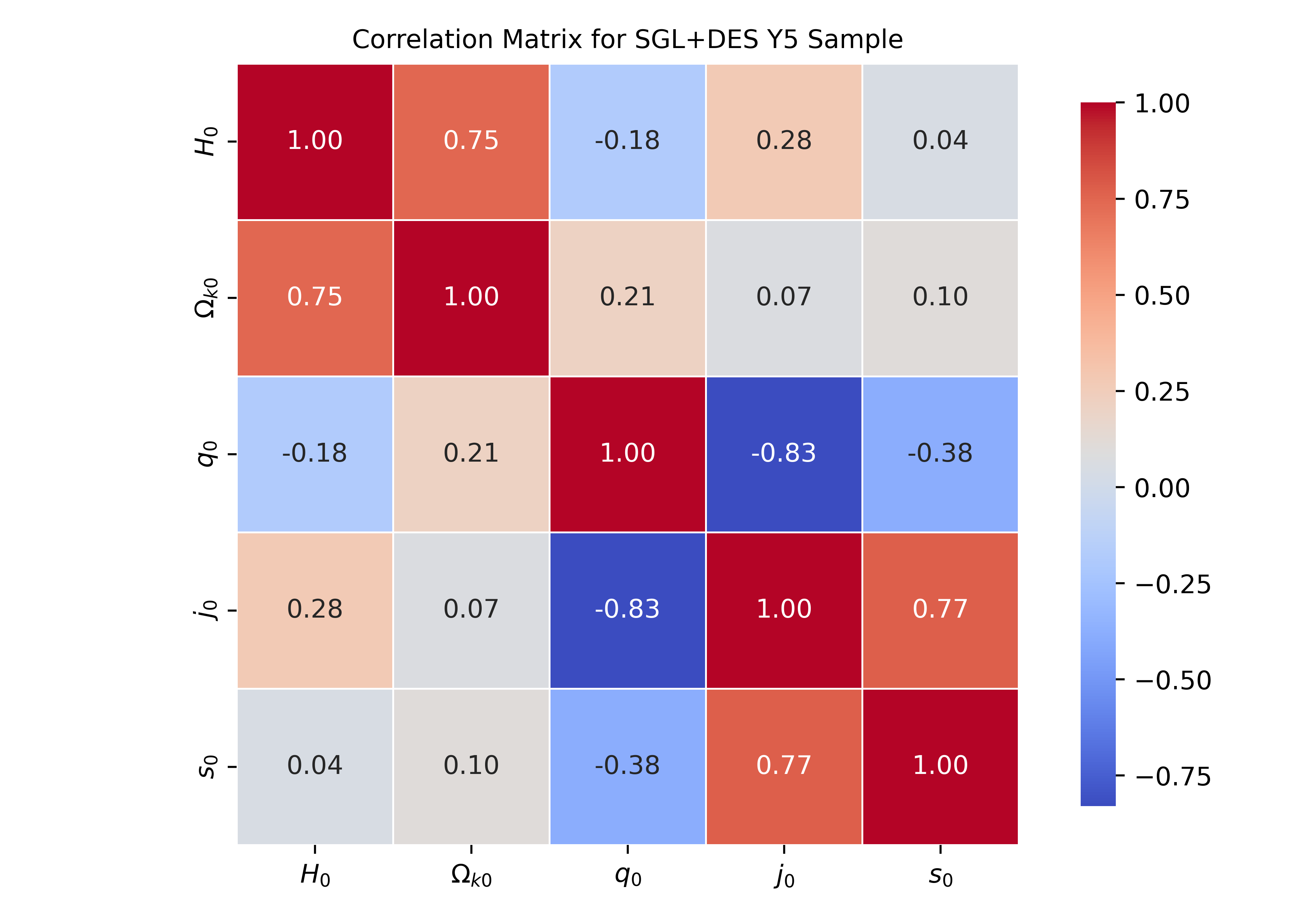}
    \label{fig_corr_union3}}
    \hspace{0.05\linewidth}
    \subfloat[Correlation matrix: SGL+ DESY5+DESI-DR2]
    {\includegraphics[width=0.45\linewidth]{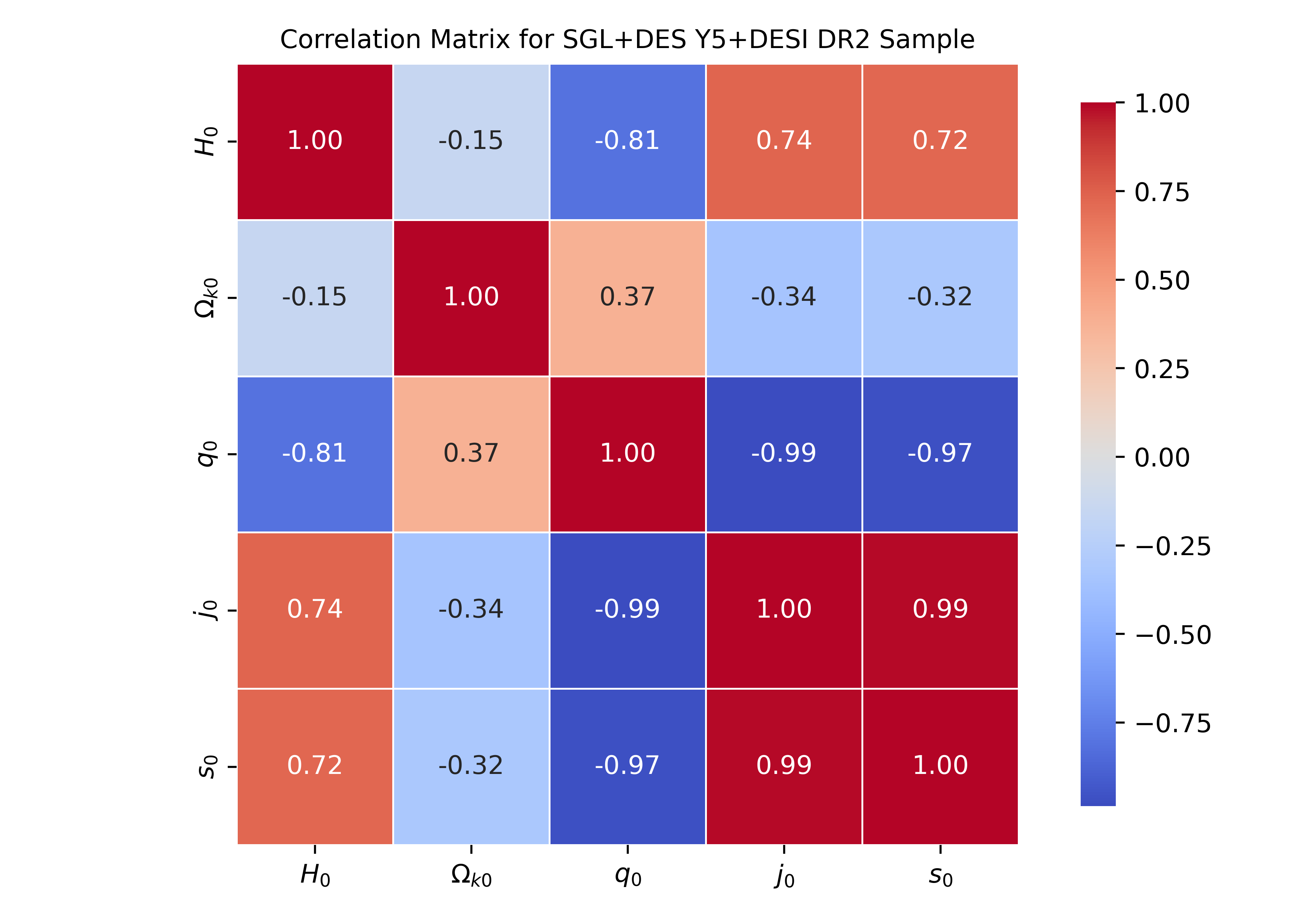}
    \label{fig_corr_desy5_desi}}

       \caption{(a) Posterior contours for SGL+DESY5 and SGL+DESY5+DESI-DR2 datasets; (b)–(c) corresponding correlation matrices for each dataset combination.}

    \label{fig_combined_contour_corr_desy5}
\end{figure}
  
Figures \ref{fig_combined_contour_corr_panthp}-\ref{fig_combined_contour_corr_desy5} display the joint $68\%$ and $95\%$ confidence contours and correlation matrices for the combinations of SGL with different SNIa datasets, with and without the inclusion of DESI-DR2. In all cases, the inclusion of DESI-DR2 data consistently provides much tighter constraints on the cosmological parameters, which is clearly visible in the reduced area of the confidence contours. \\

For the SGL+PantheonPlus dataset (Figure \ref{fig_combined_contour_corr_panthp}), the contour regions are broader in the absence of DESI-DR2. {The contours for $H_0$ and $\Omega_{k0}$ show a degeneracy without DESI-DR2. This degeneracy disappears once DESI-DR2 is added, and the central values of these parameters move closer to the $\Lambda$CDM expectations.} The contours for cosmographic parameters also become narrower with DESI-DR2, and this confirms that it plays an important role in reducing uncertainties.\\            

Figure \ref{fig_combined_contour_corr_union3} shows that the Union3 dataset without DESI-DR2 indicates significant degeneracies between $\Omega_{k0}$ and $q_0$, with a noticeable shift towards positive curvature and less negative deceleration. The inclusion of DESI-DR2 resolves these ambiguities and aligns the contours more closely with the standard model predictions. Further, the constraining power of DESI-DR2 is further evidenced by the tightened contours on cosmographic parameters, which confirms its utility in reducing statistical uncertainties.\\

Figure \ref{fig_combined_contour_corr_desy5} shows the DESY5 dataset follows a similar pattern. Without DESI-DR2, the constraints remain weak and prone to systematic uncertainties. With DESI-DR2, the contours for all parameters, especially $\Omega_{k0}$ and $q_0$, contract substantially, showing a stronger alignment with flat geometry and accelerated expansion.\\

Overall, the contour plots clearly demonstrate the critical role of DESI-DR2 in improving the analysis. The inclusion of DESI-DR2 data not only shifts the central values for $H_0$, $\Omega_{k0}$, $q_0$, $j_0$, and $s_0$ closer to the predictions of the $\Lambda$CDM model but also leads to a substantial reduction in uncertainties by a factor of approximately 2–3 in most cases. These improvements highlight the powerful contribution of DESI-DR2 towards achieving more precise and reliable cosmological constraints. \\


\section{Discussion and Conclusions}\label{sec_disc_conc}     
In this paper, we present a cosmographic analysis using a combination of strong gravitational lensing time-delay distance measurements and Type Ia supernova observations, with and without the inclusion of the DESI-DR2 dataset. Our aim is to constrain key cosmological parameters such as the Hubble constant ($H_0$), the curvature parameter ($\Omega_{k0}$), the deceleration parameter ($q_0$), the jerk parameter ($j_0$), and the snap parameter ($s_0$). We systematically show how the combination of these observational probes, particularly the DESI-DR2, improves parameter estimation and reduces uncertainties. The analysis follows Bayesian inference using Markov Chain Monte Carlo (MCMC) methods, and we marginalize over nuisance parameters where necessary. We compare the results with the standard $\Lambda$CDM model, and we critically evaluate the role of different dataset combinations using statistical diagnostics such as confidence contours and deviation tension matrices. \\

In previous studies, cosmographic analyses primarily focused  on observations derived from the distance ladder \cite{2010JCAP...03..005V,2013MPLA...2850080L,2020ApJ...900...70R}, while other works investigated the distance sum rule (DSR) as a method to constrain cosmic curvature using various observational datasets \cite{2019MNRAS.483.1104Q, 2019PhRvL.123w1101C,2023MNRAS.521.4963D,2021MNRAS.503.2179Q,2019PhRvD..99h3514L}. Some researchers combined DSR with strong gravitational lensing (SGL) data, but their distance calibration relies exclusively on distance ladder measurements \cite{2021PhRvD.103f3511K,2019PhRvL.123w1101C,2023MNRAS.521.4963D}. \textit{In this study, we introduce for the first time a comprehensive framework that integrates the distance sum rule with cosmographic parameters and strong gravitational lensing time-delay distance measurements to explore the geometry of the universe. Additionally, we incorporate the DESI-DR2 dataset to evaluate its impact on parameter estimation and to strengthen the constraints, thereby demonstrating its significant contribution to improving cosmological analyses.} This methodology represents a novel and independent approach that extends beyond earlier efforts by offering new ways to test cosmological models and understand the expansion history of the universe.\\

\begin{figure}[!ht]
    \centering
    \includegraphics[width=1\linewidth]{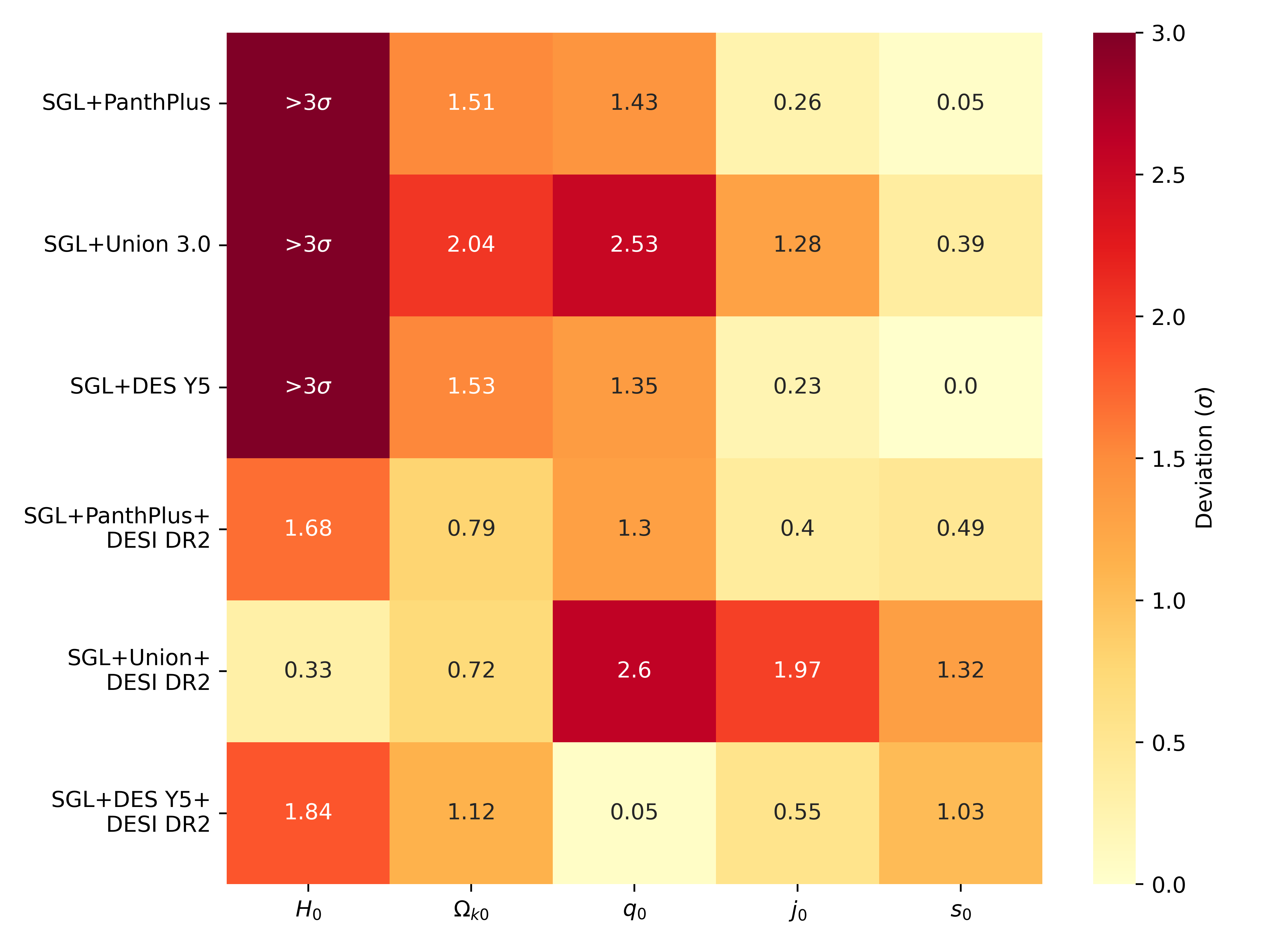}   
    \caption{Tension matrix showing the statistical deviation (in units of $\sigma$) of cosmographic parameters ---  deceleration parameter ($q_0$), jerk parameter ($j_0$), and snap parameter ($s_0$) --- with respect to the reference $\Lambda$CDM values ($H_0$ from Planck, $\Omega_{k0}=0$, $q_0 = -0.55$, $j_0 = 1$, $s_0 = -0.35$) across seven different combinations of observational datasets. }                     
    \label{fig:enter-label}
\end{figure}

Our main conclusions are listed below:                
\begin{itemize}
    \item \textbf{Constraints without DESI-DR2:} When using SGL data combined with individual SNIa datasets such as PantheonPlus, Union3, and DESY5, the parameter constraints remain relatively weak. {The estimated values of $H_0$ are consistent with the local value obtained from SH0ES survey \cite{2022ApJ...934L...7R}. The curvature parameter $\Omega_{k0}$ shows small deviations from flat geometry and indicates slight curvature effects that remain within statistical uncertainties.} The deceleration parameter $q_0$ tends to be less negative, which indicates a lower cosmic acceleration compared to the expectation from $\Lambda$CDM ($q_0 \approx -0.55$). The constraints on the jerk parameter $j_0$ and the snap parameter $s_0$ are broad, with $s_0$ being particularly weak and carrying large uncertainties. Despite these limitations, the inclusion of time-delay samples plays a critical role by establishing the cosmic distance scale. Their contribution is visible in stabilizing the parameter estimation but remains limited by the statistical power of the SNIa datasets alone.

    \item \textbf{Constraints with DESI-DR2:} The addition of DESI-DR2 data dramatically enhances the precision of the parameter estimates. The values of $H_0$ align closely with Planck’s measurements, and the uncertainties decrease by factors of 3–4 across the datasets. The curvature parameter $\Omega_{k0}$ tightens around zero at the 68\% confidence level, reinforcing the assumption of spatial flatness. The deceleration parameter $q_0$ becomes more consistent with $\Lambda$CDM predictions, and the jerk parameter $j_0$ converges towards the expected value of unity. Notably, the snap parameter $s_0$ also benefits from the additional data, as it produces significantly tighter constraints compared to the analysis without DESI-DR2. The time-delay samples provide essential information that strengthens the distance measurements and substantially improves the reliability of the cosmographic analysis.

   { \item \textbf{Curvature sign flip:} The SGL+SNIa results show a mildly positive curvature, consistent with a slightly open universe, although the deviation remains within statistical uncertainties. When the DESI-DR2 dataset is added to form a joint SGL+SNIa+DESI analysis, the curvature parameter becomes negative, indicating a mildly closed universe.  This shift occurs because DESI-DR2 provides precise baryon acoustic oscillation distance measurements, which constrain the overall geometry of the universe more tightly and change the relative weight of low-redshift SNIa distances and high-redshift SGL measurements. This sign flip indicates the sensitivity of cosmic curvature constraints to the inclusion of high-precision BAO measurements from DESI.}


    \item \textbf{Comparison with $\Lambda$CDM:} The results obtained without DESI-DR2 reveal some deviations from $\Lambda$CDM, particularly in $q_0$ and $\Omega_{k0}$, although these differences remain statistically consistent with the standard model within the error bounds. {The inclusion of DESI-DR2 data, however, makes the constraints on the cosmographic parameters much tighter and reduces the size of the error bars for $q_0$, $j_0$, and $s_0$. The tension matrix shown in Figure~\ref{fig:enter-label} further confirms this trend. The figure presents the statistical deviation in units of $\sigma$ for $q_0$, $j_0$, and $s_0$ across different combinations of observational datasets. It clearly indicates that uncertainties in cosmographic parameters appear larger in analyses without DESI-DR2, particularly for $s_0$, where the constraints remain weak and less reliable. The inclusion of DESI-DR2 leads to a significant reduction in the uncertainties and suggests that the tensions observed in earlier analyses may arise from the inherent limitations of the time-delay and SNIa datasets rather than from fundamental flaws in the $\Lambda$CDM model.}

    \item \textbf{Contour Analysis and Degeneracies:} The contour analysis based on the joint likelihoods shows that, without DESI-DR2, the parameter spaces display broad contour regions. The regions between $H_0$ and $\Omega_{k0}$, and between $q_0$ and $s_0$, remain extended, limiting the precision of the inferred constraints. The inclusion of DESI-DR2 data narrows these contours and reduces the degeneracy between $H_0$ and $\Omega_{k0}$, leading to a more compact and well-defined parameter region. The correlation matrices, shown in Figures~ \ref{fig_combined_contour_corr_panthp}-\ref{fig_combined_contour_corr_desy5}, clearly illustrate this trend. In all cases, DESI-DR2 reduces the degeneracy between $H_0$ and $\Omega_{k0}$, while the correlations among the cosmographic parameters increase. The cross-correlation between the cosmological parameters ($H_0$, $\Omega_{k0}$) and the cosmographic parameters ($q_0$, $s_0$) also shows a modest rise when DESI-DR2 is included. This pattern indicates that the combined dataset constrains the overall parameter space more coherently. The time-delay data from SGL remains critical for refining parameters, and the tightened contours along this axis confirm its impact. Overall, the contour plots and correlation matrices together provide a consistent view of the improvement achieved through the inclusion of DESI-DR2.
    

\end{itemize}

{This study shows that strong lensing time-delays combined with supernova data can probe spatial curvature. The role of DESI-DR2 is decisive: without DESI-DR2 the preferred results indicates an open universe, while the addition of DESI-DR2 flips the sign and favors a closed universe. This shift illustrates the degeneracy between curvature and late-time expansion parameters and highlights the necessity of DESI-DR2 for better results. A companion paper (\textbf{Part II}) turns to lensing distance ratios, which address curvature differently. Time-delay analyses treat $H_0$ as a free parameter, while the distance ratio analysis does not, and this distinction allows the two approaches to complement one another.}\\

While our analysis has achieved meaningful improvements in parameter estimation, it is important to acknowledge potential systematic effects and limitations. The SGL time-delay measurements, though powerful, depend on precise modeling of lens mass profiles, line-of-sight structures, and environmental effects, any of which can introduce biases if not properly addressed. Similarly, supernova samples are subject to calibration uncertainties, selection effects, and possible redshift-dependent biases that may influence distance estimates. The large uncertainties observed in higher-order parameters such as $j_0$ and $s_0$ indicate that current datasets do not fully constrain cosmic expansion beyond the second derivative. These challenges underline the need for refined observational strategies and data processing methods that reduce systematic errors and improve the reliability of cosmographic analyses.\\

The expansion of observational efforts through upcoming surveys will transform the prospects for cosmographic studies. Ongoing programs such as the Dark Energy Survey \cite{2018MNRAS.481.1041T} and the Hyper Suprime-Cam Survey \cite{2017MNRAS.465.2411M} already contribute valuable datasets, while next-generation surveys including the Vera C. Rubin Observatory’s LSST \cite{2010MNRAS.405.2579O}, the Euclid mission, and the Nancy Grace Roman Space Telescope \cite{2018PhR...778....1B, 2018A&A...614A.103P} will vastly increase the discovery of lensed quasars and other transient sources with precisely measured time-delays. Space-based telescopes such as the Hubble Space Telescope and ground-based adaptive optics facilities will allow for accurate modeling of stellar kinematics in lens galaxies, further reducing uncertainties in mass profiles. Additionally, strongly lensed supernovae, as anticipated in recent studies \cite{2022ChPhL..39k9801L}, will serve as new probes of cosmic expansion, while improved Type Ia supernova datasets will densely map the expansion history across a wide range of redshifts. These developments will not only improve measurements of the Hubble constant and parameters like $\Omega_{k0}$, $\gamma_\mathrm{PPN}$ but also offer stronger tests of general relativity and alternative cosmological models, thereby setting the stage for a deeper understanding of the universe’s geometry and evolution.\\     

There are several promising directions for extending this work. The inclusion of additional observational probes such as baryon acoustic oscillations, cosmic chronometers, and weak lensing measurements could enhance the robustness of cosmographic constraints. A thorough approach to modeling systematics, combined with cross-validation among independent datasets, will reduce biases and confirm the reliability of results. The exploration of extended cosmographic frameworks and alternative gravity models could also provide deeper understanding of nature of dark energy and cosmic acceleration. Finally, the application of advanced statistical tools and machine learning techniques may open new avenues for efficiently handling complex parameter spaces and extracting meaningful constraints from high-precision datasets.\\

In summary, our analysis demonstrates that combining time-delay measurements from strong lensing systems with supernova datasets, especially with the inclusion of DESI-DR2, produces robust constraints on cosmological parameters and improves consistency with the standard $\Lambda$CDM model. We believe that the results presented here will help  guide future observational strategies and cosmological investigations aimed at refining our understanding of the universe’s expansion history and geometry.

\acknowledgments  
Darshan Kumar is supported by the Startup Research Fund of the Henan Academy of Sciences, China under Grant number 241841219. One of the authors (Deepak Jain) thanks Inter-University Centre for Astronomy and Astrophysics (IUCAA), Pune (India) for the hospitality provided under the associateship programme where part of the work was done. In this work the figures were created with \textbf{{\texttt{GetDist}}}~\cite{2019arXiv191013970L}, \textbf{ {\texttt{numpy}}}~\cite{numpy}  and \textbf{{\texttt{matplotlib}}}~\cite{matplotlib} Python modules and to estimate parameters we used the publicly available MCMC algorithm  \textbf{ {\texttt{emcee}}} \citep{emcee}. 

   
   
\bibliographystyle{JHEP}
\bibliography{main.bib}
   

\end{document}